\documentclass[journal]{vgtc}                

\ifpdf
  \pdfoutput=1\relax                   
  \usepackage{graphicx}                
  \DeclareGraphicsExtensions{.pdf,.png,.jpg,.jpeg} 
\else
  \usepackage{graphicx}                
  \DeclareGraphicsExtensions{.eps}     
\fi%

\graphicspath{{figures/}{pictures/}{images/}{./}} 

\usepackage{microtype}                 
\PassOptionsToPackage{warn}{textcomp}  
\usepackage{textcomp}                  
\usepackage{mathptmx}                  
\usepackage{times}                     
\usepackage{cite}                      
\usepackage{tabu}                      
\usepackage{booktabs}                  

\usepackage{enumitem}
\usepackage{outlines}
\usepackage{floatrow}
\usepackage{multirow}
\usepackage{flushend}

\onlineid{0}

\vgtccategory{Research}
\vgtcpapertype{evaluation}

\title{Measuring Effects of Spatial Visualization and Domain on Visualization Task Performance: A Comparative Study \vspace{-0.1in}}

\author{Sara Tandon, Alfie Abdul-Rahman, and Rita Borgo}
\authorfooter{
\item 
All the listed authors are affiliated with King's College London. E-mail: 
\{sara.tandon $|$ alfie.abdulrahman $|$ rita.borgo\}@kcl.ac.uk.
}

\shortauthortitle{Tandon \MakeLowercase{\textit{et al.}}: Measuring Effects of Spatial Visualization and Domain on Visualization Task Performance}

\abstract{Understanding your audience is foundational to creating high impact visualization designs. However, individual differences and cognitive abilities also influence interactions with information visualization. Differing user needs and abilities suggest that an individual’s background could influence cognitive performance and interactions with visuals in a systematic way. This study builds on current research in domain-specific visualization and cognition to address if domain and spatial visualization ability combine to affect performance on information visualization tasks. We measure spatial visualization and visual task performance between those with tertiary education and professional profile in business, law \& political science, and math \& computer science. We conducted an online study with 90 participants using an established psychometric test to assess spatial visualization ability, and bar chart layouts rotated along Cartesian and polar coordinates to assess performance on spatially rotated data. Accuracy and response times varied with domain across chart types and task difficulty. We found that accuracy and time correlate with spatial visualization level, and education in math \& computer science can indicate higher spatial visualization. Additionally, we found distinct motivations can affect performance in that higher motivation could contribute to increased levels of accuracy. Our findings indicate discipline not only affects user needs and interactions with data visualization, but also cognitive traits. Our results can advance inclusive practices in visualization design and add to knowledge in domain-specific visual research that can empower designers across disciplines to create effective visualizations.%
} 


\CCScatlist{ 
 \CCScat{K.6.1}{Management of Computing and Information Systems}{Project and People Management}{Life Cycle};
 \CCScat{K.7.m}{The Computing Profession}{Miscellaneous}{Ethics}
}

\teaser{
  \centering
  \includegraphics[width=\textwidth]{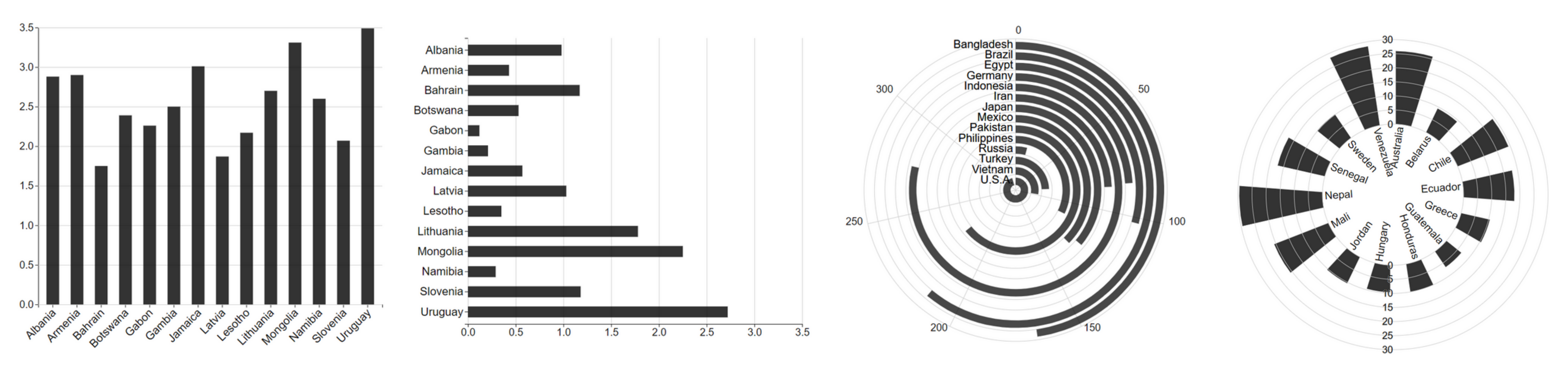}\vspace{-0.15in}
  \caption{ Stimuli created for this study displaying a bar chart rotated along Cartesian and radial axes: Vertical, Horizontal, Radial, and Circular Bar Plot.}
  \label{fig:teaser}
}

\vgtcinsertpkg
\usepackage{subcaption}

\begin{document}


\firstsection{Introduction}

\maketitle

Data shapes modern society and affects every aspect of our lives: political, social, and physical. Information visualization is just as prevalent -- it lives in magazines, news outlets, scientific papers, blogs, and countless online platforms to facilitate engagement with data in a way text alone cannot. The data visualization community has long explored how to design universally accessible and engaging visualizations that empower communities to understand and reason with data~\cite{peck_data_2019}. Information visualization is not simply a matter of reading data properties encoded in visual form -- there are complex cognitive activities at work that are influenced by visual structure~\cite{ziemkiewicz_perceptions_2009}. Visualization research has followed several paths toward understanding users and their needs including understanding how cognitive abilities affect perception and how disciplines differ in terms of data and tasks~\cite{ottley_improving_2016, velez_understanding_2005, ziemkiewicz_perceptions_2009,KirbyCollab,DSM2012}. Education psychology has explored the connections between education background, performance, and the cognitive ability of spatial visualization, noting these are not easily separated in individuals~\cite{Atit2020ExaminingTR,Wai2009SpatialAF}. Recently, work has been done in visualization toward integrating these tracks of research to move away from universal guidelines for the monolithic ``user'' and towards effective design that considers the differing cognitive traits of target disciplines~\cite{Hall2022ProfessionalDA}. 

We build on this past research to explore and investigate the cognitive trait of spatial visualization and its interaction within and across not only professional domains but also educational backgrounds. We focus our study on the disciplines of business, law \& political science, and math \& computer science; the use of visualization is core to these disciplines, from decision making to policy development to data analysis. In this paper, we present past research related to spatial visualization in psychology and visualization design that led us to create hypotheses around how our target domains would perform spatial visualization assessments and relevant visual tasks. We pose a study methodology that links spatial visualization abilities to visual task paradigms that evoke the same cognitive functions. Finally, we offer a detailed statistical analysis demonstrating that visualization task performance (accuracy and response times) varies with spatial visualization and discipline. Additionally, we present how domain differences in motivations around use of data visualization might affect performance on visual tasks. 

The results we collected add to evidence that spatial visualization and domain interact to affect visual task performance and should be considered as combined factors that can influence design choices; visualization design should consider not only the needs of varying domains, but the abilities of individuals in those domains~\cite{Hall2022ProfessionalDA}. This paper contributes toward cataloguing visualization performance differences amongst spatial visualization levels and domains that can lead to empowering individuals across disciplines to make memorable references and inferences effectively and efficiently using visualizations designed for them.

\section{Related Work}

Our study builds on research in spatial visualization, education psychology, and information visualization. It bridges visualization research in user cognition and domain needs/expertise to inform design as a holistic process. We draw on methods from psychology and visualization to investigate the relationships between spatial visualization, domain, and visualization comprehension. 

\subsection{Spatial Visualization}
Spatial ability is a cognitive ability referring to skills involved with retrieval, retention, and transformation of visual information~\cite{Michael1957TheDO, kimura_sex_2000}. It impacts a viewer's use of graphics as it influences the processing of spatial relationships between graphical elements and/or their cognitive resource allocation~\cite{vanderplas_spatial_2016}. Spatial visualization is a specific aspect of spatial abilities that involves the ability to remain unconfused by varying orientations or rotational positions in which a spatial pattern may be presented, thus it is elicited most in tests that involve manipulation or transformation of a visual stimulus~\cite{Michael1957TheDO, kimura_sex_2000}. Spatial visualization allows an individual to compare different encodings of visualizations quickly and accurately~\cite{Michael1957TheDO}, and has been connected to spatial scaling, figure construction, magnitude sense, numerical estimation, subdivisions of charts, and even textual spatial analogies~\cite{Young2018TheCB,MIX2012197}. In all, high spatial visualization implies the ability to manipulate and change internal representations of visualizations, leading to measurable outcomes when interacting with external visualizations~\cite{liu_mental_2010, cohen_individual_2007, hegarty_diagrams_2004}.

\subsection{Education Psychology and Spatial Visualization}
Spatial abilities, including spatial visualization, have often been tested in educational settings as an indicator of increased performance in STEM subjects, especially in performance in mathematical subjects~\cite{Salthouse1990AgeAE, MIX2012197, jones_spatial_2008, Atit2020ExaminingTR, Shea2001ImportanceOA, Young2018TheCB, FERGUSON20151}. Increased spatial visualization ability has been specifically tied to increased performance in students of geology and architecture as well~\cite{Orionearthscience, Salthouse1990AgeAE}. Additionally, substantial spatial visualization deficits have been identified as the primary component in certain learning disorders wherein students struggle to use visual and graphic representations of data~\cite{broitman_nvld_2020}. 

Research has also been done into spatial visualization of students and professional across a range of domains~\cite{Burnett1980EffectsOA, Shea2001ImportanceOA, Wai2009SpatialAF}. Burnett and Lane~\cite{Burnett1980EffectsOA} tested college students in their first semester and again after two years of study in various subjects. They found that spatial visualization performance improved with advance in education across all subjects, but by a lower magnitude for humanities and social science students than math and physical science students. Shea et al.~\cite{Shea2001ImportanceOA} and Wai et al.~\cite{Wai2009SpatialAF} conducted comprehensive studies assessing spatial abilities across domains, levels of education, and occupation. Both found students of humanities, education, law, business, and medicine to have lower spatial abilities than math/computer science, physical science, and engineering students. This shows that specific domains and duration of experience could influence individual ability to use information visualization. Evidence that it is difficult to separate spatial visualization abilities from education and professional choices could empower designers to create impactful visuals for specific communities.

\subsection{Spatial Visualization in Information Visualization}
A rich body of literature exists investigating spatial abilities in the context of information visualization
\cite{ottley_improving_2016, velez_understanding_2005, ziemkiewicz_perceptions_2009, kim_explaining_2017, vicente_assaying_1987, vanderplas_spatial_2016, wenhong_user_2019, Hall2022ProfessionalDA}.
Research has demonstrated that high spatial visualization specifically, can correlate with higher recall, understanding, and increased task performance using data visualization~\cite{kim_explaining_2017, ottley_improving_2016}. 

Kim et al.~\cite{kim_explaining_2017} observed that for both parallel coordinate plots and written descriptions of data, participants with high spatial visualization had higher accuracy when recalling values. Vicente et al.~\cite{vicente_assaying_1987} found low spatial ability corresponded with inferior performance on retrieval tasks within visual file structures: low spatial individuals were two times slower and more likely to get lost in the structure than high spatial individuals. Kellen~\cite{kellen_effects_2012} and Ottley et al.~\cite{ottley_improving_2016} found that high spatial individuals had significantly increased accuracy in their performance on solving conditional probability problems when aided by a visualization. On their study using line-up tests, VanderPlas and Hofmann~\cite{vanderplas_spatial_2016} found that performance of undergraduate students correlated with visual ability and whether a student was a STEM major. Studies also indicated that speed is often related to understanding~\cite{stanney_information_1995, velez_understanding_2005} and was sometimes the bigger indicator of lower cognitive abilities compared to accuracy~\cite{wenhong_user_2019, conati_evaluating_2014}. Wenhong~\cite{wenhong_user_2019} found that high spatial ability led to reduced response times and higher accuracy with graphics and table visualizations, and interestingly that those with low spatial ability still chose graphic visualizations for tasks. Wenhong noted that often individuals are not aware of their own cognitive style, but still prefer visual representations of data. These findings indicate space for accessible design in information visualization for low spatial individuals~\cite{Wu2021UnderstandingDA}.

While cognitive abilities are thoroughly studied in visualization design, educational or professional domain is often overlooked as a relevant influence and not reported on \cite{kim_explaining_2017,kellen_effects_2012,vicente_assaying_1987, ottley_improving_2016, velez_understanding_2005, stanney_information_1995} or referenced, but not analyzed~\cite{conati_evaluating_2014, wenhong_user_2019}.  However, there is also rich research in domain-focused participatory methodologies that imply domain can also impact design. 

\subsection{Designing Visualization for Domain}

Literature on visualization design highlights interaction with, and discovery of, needs for different domains to inform the design process, but they do not explicitly consider differing cognitive traits of the domains~\cite{DSC:7Scen2018, DSM2012, Hall2020DesignBI, BertiniBridge2015, KirbyCollab, NestedModel2009}. The design study methodology proposed in \cite{DSM2012} emphasizes the fundamental need to design with and for the needs of specific domains. Distinct from discovering practices and requirements of specific domains, it also advocates for designers to be aware of literature pertaining to relevant design problems, but not specifically to cognitive and perceptive knowledge, which are themselves distinct from visual task needs of domain experts. Other methods like participatory design, design by immersion, and domain liaisons~\cite{Hall2020DesignBI, BertiniBridge2015} potentially have benefits due to implicitly integrating design reflective of cognitive capabilities, without the need to explicitly rely on any empirical, systematic analysis of cognitive abilities. 

Kirby \& Meyer~\cite{KirbyCollab} and Munzner~\cite{NestedModel2009} further emphasize collaboration with domain experts and employment of applied psychology methodologies to measure cognition and perception, but do not touch upon capitalizing on the cognitive features of domains in guiding design. Peck et al.'s multidimensional model of individual differences in HCI~\cite{Peck3DModel} separates individual experiences/bias from cognitive abilities in visualization design:~\cite{Hall2022ProfessionalDA} notes that treating these as orthogonal traits implies an individual's background and experiences do not influence their cognitive traits and vice versa, which could lead to missed opportunities and ineffective design choices. 

Hall et al.~\cite{Hall2022ProfessionalDA} made an initial exploration into how spatial visualization differs between professional domains (Education, Chemistry, and Computer Science) and relates to visualization processing. They confirmed spatial visualization differences amongst disciplines and that domain and spatial visualization together influenced performance and qualitative strategies on tasks related to 2D to 3D representations of isocontour plots and scatter plots. Their findings imply differences in terms of task performance and spatial visualization between domains and that these differences will endure with design implications for visualization. Research demonstrates there are ties between spatial visualization, domain, and visualization performance - we endeavor to further findings into how cognitive differences between domains relate to visualization performance to affect design decisions and empower designers to create more relevant and cognitively appropriate visuals.

\begin{figure*}[ht]
 \centering 
 \includegraphics[width=\textwidth]{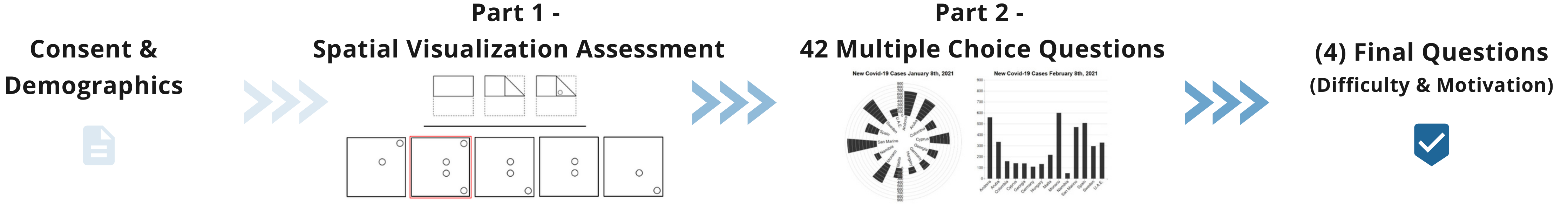}
 \caption{Activity sequence of the study. In Part 1, the paper is folded and punched above the line. Participants are given four multiple-choice responses to choose between, the selected choice is highlighted and participants can move forward. Scores are the number answered correctly out of 10 in 3 minutes. \vspace{-0.2in}}

\label{fig:studyorg}
\end{figure*}

\section{Motivation \& Hypothesis}\label{sec:Hypothesis}
Drawing on the outcomes and identified gaps from previous work, we sought out to study spatial visualization across domains and their combined influences on visualization interaction. We ensured alignment between profession and education background as research demonstrates increased spatial visualization depending on education domain~\cite{Burnett1980EffectsOA}. If occupation does not align with educational background, the abilities gained in education should be considered. 

We chose to study those with professional profile and education background in Law and Political Science (LPS), Business, and Math and Computer Science (MCS). Differences between MCS and other disciplines have implications on visualization research findings as they are often ``standard participants'' for many studies on task performance and design choices; a considerable amount of information visualization is also created and studied by those in MCS related fields~\cite{chen_individual_2000, chen_spatial_1997, vanderplas_spatial_2016}. We chose Business as there is a wealth of literature exploring data visualization for business intelligence and business decisions, which often cites design and cognition research~\cite{few2007data,zheng2017data, tegarden1999business, luo2019user}. However, we are unaware of any comparative studies between MCS and Business domain cognitive abilities applied to data visualization task performance. We chose LPS due to the relative lack of research investigating data visualization and law \& politics, despite the imperative use of data in law \& politics~\cite{zinovyev2010data, henshaw2018data}. However, given many governments have scientific advisory and consulting committees, and politicians make health and governance related decisions based on this advice~\cite{Colmane006928}, it may be worth exploring any differences in visual cognition between those creating and consuming visualizations for political decisions. 

Additionally, these domains display varying levels of spatial visualization ability with MCS exhibiting higher than average abilities, and Business and LPS below average~\cite{Wai2009SpatialAF}. Given that spatial visualization affects interactions with visualization, there may be measurable differences in how these domains interact with data visualizations given their discipline. Our study aims to explore if spatial visualization ability differs across domain, how spatial visualization level affects interactions with data visualization, and if domain interacts with spatial visualization to affect performance in a quantitatively significant way. We explore these questions by investigating the following hypotheses.


\begin{description}
  
\item[H1:] \textbf{Spatial visualization will differ across domains.} Research has demonstrated systematic differences in spatial visualization abilities between domains~\cite{Shea2001ImportanceOA, Wai2009SpatialAF}. Drawing from these findings, we anticipate MCS will have the highest spatial visualization ability and LPS will have the lowest. As spatial visualization affects visualization interaction, these systematic differences between domains may impact visual reasoning in a discriminating way. 

\item[H2:] \textbf{Performance (accuracy and response time) will correlate with spatial visualization level.} Prior work in information visualization shows accuracy and timing of visualization tasks vary with spatial visualization abilities~\cite{ottley_improving_2016, velez_understanding_2005, ziemkiewicz_perceptions_2009, kim_explaining_2017, vicente_assaying_1987, vanderplas_spatial_2016, wenhong_user_2019, Hall2022ProfessionalDA}. High spatial individuals tend to have higher accuracy and reduced response times. We expect to see comparable results, emphasizing the role of spatial visualization in performance and use of visualizations. 

\item[H3:] \textbf{Performance (accuracy and response time) will correlate with domain such that average performance will vary between disciplines.} Hall et al.'s research~\cite{Hall2022ProfessionalDA} partially confirms this hypothesis between Chemistry, CS, and Education. They found similar performance and spatial visualization between CS and Education, but observed an overall time/error trade-off between the disciplines regardless of spatial visualization levels. However, they found correlations between spatial visualization and task performance across all disciplines, implying that spatial visualization is an important underlying source of variations. We anticipate marked differences between MCS, Business, and LPS given the three domains varying levels of interaction, creation, and consumption of data visualization~\cite{Friendly2008ABH,zheng2017data,zinovyev2010data}, with MCS performing the best and LPS the worst following their spatial visualization levels. Confirming this would substantiate the work of Hall et al., that effective visualizations should consider not only the needs of a discipline, but the abilities of the individuals within domains.

\end{description}

\section{Methodology}

A two-part online study was created to address the research questions and hypotheses (see Fig. \ref{fig:studyorg}). Part 1 of the study consisted of a brief spatial visualization psychometric test from The Kit of Factor Referenced Cognitive Tests~\cite{ekstrom_manual_1976} known as a paper folding or punch test (see Fig. \ref{fig:studyorg}). Part 2 consisted of stimuli and questions asking individuals to draw conclusions from the visualizations. Inspired by current methodology in vision science~\cite{Elliott2021ADS}, we began by evaluating paradigms that extend from psychological assessments of spatial visualization to visualization evaluation. Principles and studies in psychology dictate spatial visualization relates directly to ability to quickly and accurately compare encodings and layouts of a given visualization along with value estimation, magnitude sense, and subdivisions of charts~\cite{Michael1957TheDO, Young2018TheCB}. Working from these principles, we chose \textit{visual search} as our base paradigm~\cite{Elliott2021ADS}, utilizing data from two charts rotated along Cartesian and polar coordinates. We provide a detailed discussion of study structure and stimuli design (see supplementary material (SM) for comprehensive stimuli set).

\subsection{Recruitment}

Participants were recruited through Prolific~\cite{PalanProlific}. The participants were 18 years and older and were fluent in English. Due to the limited number of eligible participants on Prolific when filtered by education and profession, we were able to recruit a balanced sample of 30 participants in each domain with valid data, for a total of 90 participants (participant details below). Participants were paid at £7.71/hour in accordance with Prolific's fair pay policy. The average response time was $37$ minutes and $7$ seconds, resulting in an average pay of £4.78 per participant. We collected data across three days and different time slots to diversify the pool of potential participants. We collected data on Jan 28, 2022, Feb 3, 2022 and Feb 17, 2022. 

\subsection{Apparatus}
The online study was created using a Flask Web App with D3.js for chart generation. Participants went through consent and training before each part of the study and were allowed to leave the study at any time, ending their participation. A trigger warning was presented to participants before agreeing to take part in the study as the data related to the COVID-19 pandemic included case, hospitalization, death, and vaccination statistics. No identifiable data was collected and all data was stored and maintained on a private server at the authors' institution.


\begin{figure*}[t]
 \centering
    \begin{subfigure}{0.33\textwidth}
        \includegraphics[width=\textwidth]{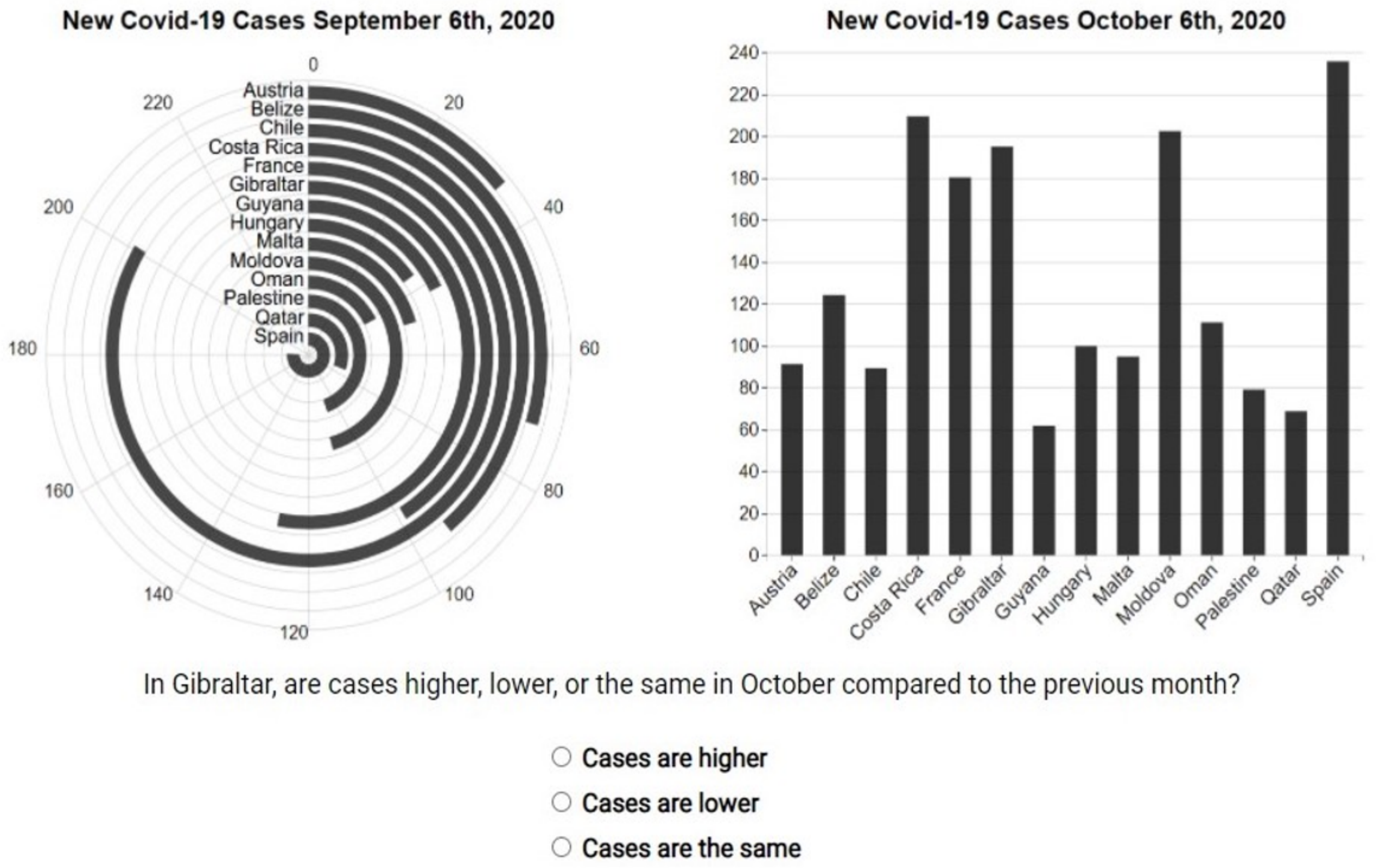}
       \caption{Easy.}
    \end{subfigure}
    \begin{subfigure}{0.33\textwidth}
        \includegraphics[width=\textwidth]{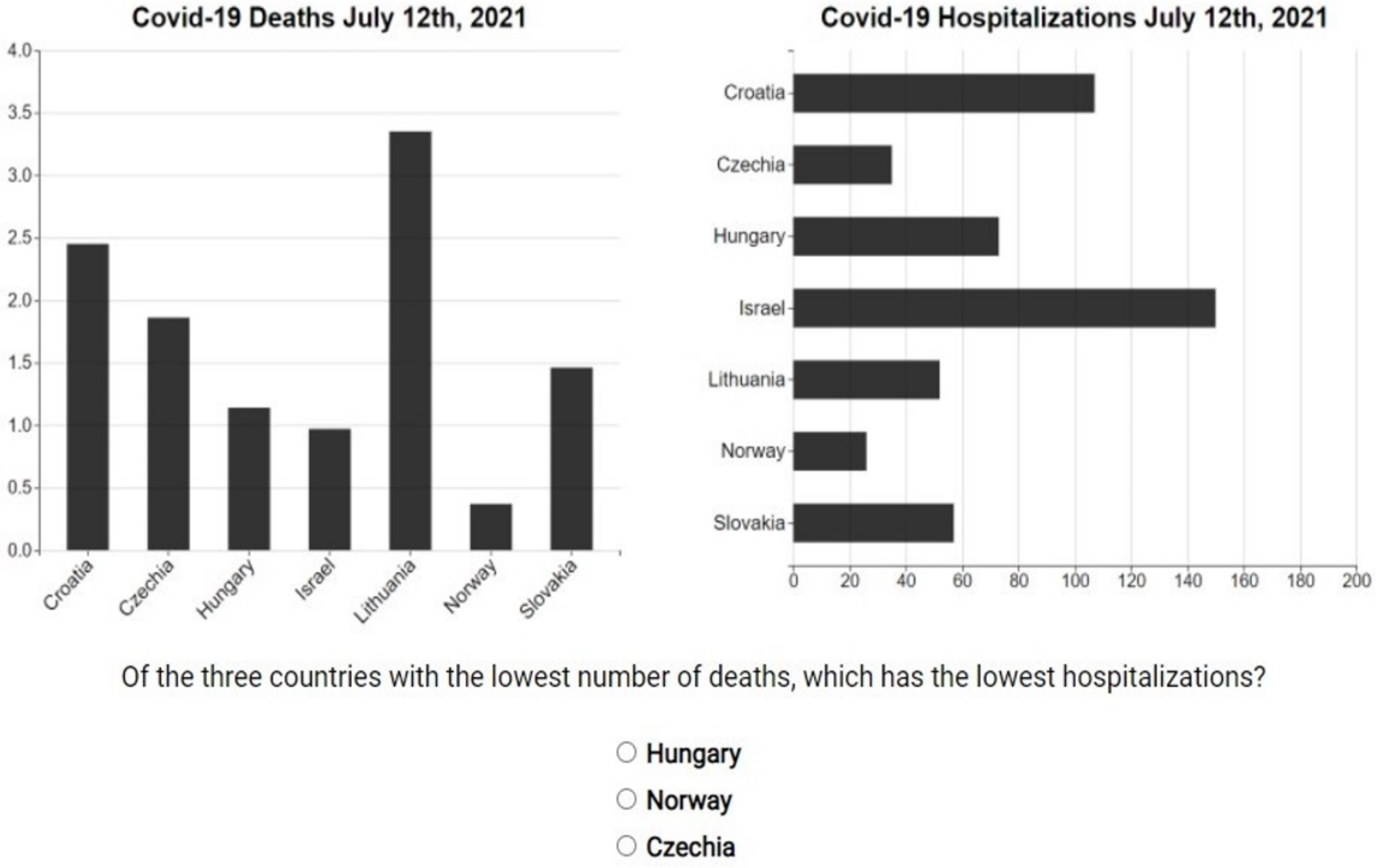}
        \caption{Medium.}
    \end{subfigure}
    \begin{subfigure}{0.33\textwidth}
        \includegraphics[width=\textwidth]{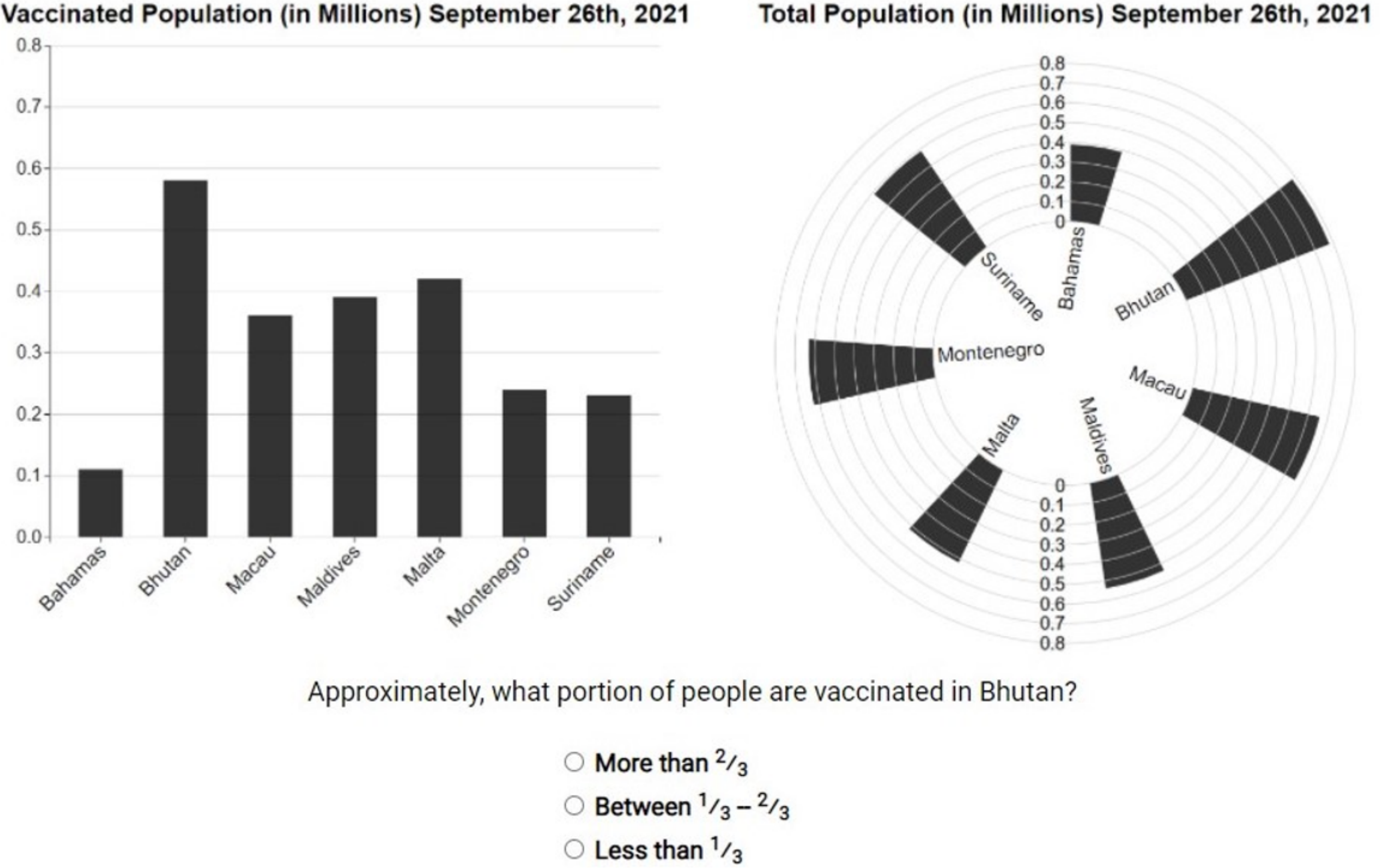}
        \caption{Hard.}
    \end{subfigure}
\label{fig:All}
 \caption{(a) Example of an Easy question with the Radial chart on left and density 14, (b) Medium question with Horizontal chart on right of density 7, and (c) Hard question with Circular Bar on the right and density 7 (refer to SM for a more comprehensive set).\vspace{-0.2in}}
 \label{fig:ThreeQs}
\end{figure*}

\subsection{Study Structure}

After consent and demographics collection, the study process consisted of two parts with an additional page of 5-point Likert scale questions, assessing perceived difficulty and personal motivations regarding data visualization (see Section \ref{sec:motiv}). Participants were asked to confirm readiness to move to the next block between each step to allow for breaks as needed. Training for Part 1 of the study came from the Kit of Factor Referenced Cognitive Tests and included a sample question that participants had to answer correctly to move forward. Active training for Part 2 consisted of 4 sample questions displaying the various charts participants would see throughout the study. 

We ran a pilot test to measure the fatigue, length, and complexity of the study. Results from the pilot suggested changes to the study workflow; it became clear that the spatial visualization assessment should be placed at the beginning to ensure accurate scores which could otherwise be hindered by fatigue from Part 2, being the longer portion of the study. Additionally, we decided to integrate a break halfway through Part 2, consisting of a catch question wherein users were encouraged to take a break, answer with a specific response, and move on when they were ready. The catch question was not analyzed (other than ensuring all respondents answered correctly for quality assurance) and was not included in response time data analysis. 

\subsubsection{Screening \& Demographics}

Following consent, participants completed a set of 7 demographic questions to collect information about gender, age, education history, profession, and countries of origin and influence. Those who self-identified on Prolific as having studied and working in one of our target areas were admitted and had to reconfirm their primary field of study and work in demographic collection. To reduce confounding factors we only selected participants where education and professional profile were aligned if they were working. No participants whose primary subject and profession failed to align with our targets were included in our analysis. Respondents could only continue through the study if participating on a laptop or desktop computer with a JavaScript enabled web browser to increase consistency of display. Prolific offers a participant score that is based on the quality of an individual's past submissions on their site: our participants had scores of 97\% or higher.

We excluded the data of 25 participants from Prolific due to either lack of consistency in primary field of study, data validation failure, or if they experienced a server interruption during the study. We gathered data until we achieved balanced samples with high quality data between our target domains.

\subsubsection{Spatial Visualization Assessment}

Following demographics, participants began the spatial visualization assessment from the Kit of Factor Referenced Cognitive Tests~\cite{ekstrom_manual_1976}, a well-established 2D psychometric assessment~\cite{Ainooson2017ACM,Michael1957TheDO} largely utilized in previous data visualization studies~\cite{kim_explaining_2017, ottley_improving_2016, MicallefAssessing12}. This assessment has the added benefit of a 3-minute time limit for participants to complete the test. This time limit contributes to limiting fatigue of participants taking an online crowdsourced test contributing to higher data quality~\cite{borgo_information_2018}. We gathered both the selected answer and the response time for each question of the assessment. The test consists of 10 questions in which an image of a paper is folded and punched; participants must then choose from 5 options what the paper will look like when unfolded (see Fig. \ref{fig:studyorg} for a sample question). The spatial visualization score is calculated as the number of correct answers out of 10. Congruent to \cite{kim_explaining_2017, ottley_improving_2016, Salthouse1990AgeAE, Downing2005TheEA} the score was centered by its mean so that participants above the mean are classified as having high spatial visualization, while participants below are classified as having low spatial visualization. 

\subsubsection{Part 2 Stimuli Design}
We manipulated three elements to analyze performance over: data density, chart type, and task difficulty. 

\paragraph{Data} We initially considered synthetic data distributions, however, opted for real-world data to favor familiarity with a data set and scenario applications with participants from a range of domains~\cite{borgo_information_2018}. The data used in this study was COVID-19 Pandemic data from a comprehensive world statistics site~\cite{worldindata}. The data reflects actual case, vaccination, hospitalization, and death statistics from March 2020-January 2022 across 95 countries. This data was chosen as the scale of the ongoing global pandemic allowed an assumption of a basic level of familiarity with the data and no need for training or expert knowledge~\cite{borgo_information_2018}. Additionally, the World Health Organization, governments, universities, and data visualization tech companies produced many COVID-19 visual dashboards for public consumption and policy recommendation during the pandemic -- many of these dashboards contained spatially rotated data of varying densities~\cite{harvardCOVID-19,msoftcovid-19,tabcovid-19,ibmcovid,whocovid}. 

The data is displayed across two levels of density: 7 and 14 points. A key manipulation for visual search is the number of objects present in the search scene (target(s) + distractor(s) = set size)~\cite{Elliott2021ADS}. We chose 7 as our first set size; 7 is known as a magic number in data visualization, as the limit of the span of absolute judgement and immediate memory sits at about 7 points~\cite{Miller1956TheMN}. This implies that moving sufficiently beyond 7 data points will increase cognitive load; thus we chose 14 as our second set size. Across stimuli, the data was ordered alphabetically by country. Conventionally, visual search studies vary the number of targets and/or distractors present over sub-conditions~\cite{Elliott2021ADS} -- we varied the two set sizes amongst the four chart types and three questions detailed below, while maintaining the same density for side-by-side charts.

\paragraph{Chart Types} The stimuli draw on elements from both psychology and previous visualization studies in spatial visualization~\cite{Hall2022ProfessionalDA, Michael1957TheDO, vicente_assaying_1987}. We formed our stimuli keeping in mind that spatial visualization directly affects ability to compare visual encodings and layouts quickly and accurately~\cite{Michael1957TheDO}, and that our spatial visualization assessment uses 2D rotated stimuli on a Euclidean plane~\cite{ekstrom_manual_1976}. We chose to display two spatially rotated charts side-by-side to assess performance as participants utilize both charts to respond to a given question. Using two spatially rotated charts side-by-side simulates the cognitive processes involved in spatial visualization assessments.
The anchoring stimuli chosen for this study was a vertical bar chart. Bar charts are often used as stimuli in cognitive ability and visualization research as they are the most ubiquitous data visualization used over the world~\cite{barral_understanding_2020} and have been shown to capture performance differences between individuals of differing abilities~\cite{Caren_highlighting, conati_eye_2014}. Additionally, spatially rotated bar charts were heavily utilized in COVID-19 pandemic data dashboards~\cite{harvardCOVID-19,msoftcovid-19,tabcovid-19,ibmcovid,whocovid}. As points on a Euclidean plane can be displayed using the Cartesian and polar coordinate systems, we use both for our stimuli~\cite{Hoffman1975AnalysisIE}. We rotated the value-axis of the bar charts in both planes. Anchoring bars on the x-axis in a Cartesian coordinate system provides a vertical bar chart with the value-axis increasing along the y-axis; anchoring bars on the y-axis results in a horizontal bar chart with the value-axis increasing along the x-axis. We then transformed to a polar coordinate plane with value-axis increasing along both $(r,\theta)$. Anchoring bars along a vertical axis with the value scale increasing with $\theta$ results in a radial bar chart. Finally, anchoring the bars at the pole with the value-axis increasing with $r$ results in a circular bar plot (see Fig. \ref{fig:teaser} for chart types). We followed conventional design principles for circular bar plots, including increasing radial distance on a linear scale given data density, and a large inner radius to reduce perception bias~\cite{healy_2022}. We note that circular bar plots have some distortion of the bar as the radial distance increases to decrease visual bias toward underestimating small bar heights~\cite{bostock_2016}. Further, previous research demonstrates that although accuracy can be inhibited in polar coordinate charts compared to Cartesian, they are still widely used and are aesthetically appealing, thus we used them as sub-conditions in this study~\cite{burch2014benefits}. We maintained the vertical bar chart throughout the questions and paired it with either itself or one of the three spatially rotated charts to allow for performance comparison between chart types (e.g., is performance between two Cartesian plots different than a Cartesian \& polar chart?). We ensured that the vertical chart appeared on the left and right of the rotated chart across densities and question types. We maintained gray-scale charts to avoid any color interference. As polar coordinate plots require chart lines for readability, we maintained similar lines on the Cartesian charts as well. We varied the four chart types amongst the two set sizes and three questions. 

\subsubsection{Part 2 Task Design}

The basic task paradigm we chose was a visual search process across two charts and densities. Visual search plays an important role in the cognitive process and is a vital element to visualization~\cite{borgo_empirical_2012}. As the stimuli consist of two side-by-side charts, the visual search process we chose was a conjunctive search: visual search involving identifying a previously requested target surrounded by distractors possessing no distinct features from the target itself~\cite{Shen2003GuidanceOE}. Conjunctive search involves search over two channels and increases with difficulty as density of distractors increase~\cite{Shen2003GuidanceOE}. We paired conjunctive search with tasks increasing in difficulty (see below) across two levels of density (as described above) to compare performance across varying levels of difficulty. We created three 3-alternative forced-choice multiple-choice questions that required conjunctive use of two chart types for response, we categorized them as Easy, Medium, and Hard. The three response options were randomized across all questions. Questions were randomized after the training section.  Fig. \ref{fig:ThreeQs} shows three examples of different levels of difficulty and density, SM provides a comprehensive overview of all the combinations.

\begin{description}
  \item[Easy Question] The easy question is a search and comparison task. It displayed two charts with case numbers from the same set of countries across two months. This question is classified as easy, as it consists of a conjunctive search for the same singular target variable across both displayed charts. Once located, participants compare the target variable across the charts i.e., participants searched for a given country in the first chart and compared case numbers of that country to the second chart to respond if cases were higher, lower, or the same as the previous month. 
  
  Easy question example: \textit{In Gibraltar, are cases higher, lower, or the same in June compared to the previous month?} Easy Responses: \textit{Cases are higher, cases are lower, cases are the same}. 
  
  \item[Medium Question] The medium question is a search and comparison task across an increased number of targets. It displayed two charts, the left with death rates, the right with hospitalization rates of the same set of countries on a given date. This question is classified as medium as it increases the variables for conjunctive search and comparison to three targets across both charts; participants searched for the three countries with the highest number of deaths in the first chart and responded with which of those countries had the highest hospitalizations from the second chart. 
  
  Medium question example: \textit{Of the three countries with the highest number of deaths, which has the highest hospitalizations?} Medium responses consisted of the correct response, one of the countries with the highest number of deaths, but not the highest hospitalizations, and one random country (see Fig. \ref{fig:ThreeQs}).  

  \item[Hard Question] The hard question is a search and comparison task that involves mathematical estimation. It displayed two charts, the left with the raw number (in millions) of vaccinated people, the right with the raw population (in millions) of the same set of countries on a given date. This question is classified as hard as after a conjunctive search for one target variable across charts, participants had to perform a mathematical computation to estimate a derived variable. Computing a derived variable is a common task in data analysis and often appears as a sub-task in other operations -- the more complex the aggregation the more difficult the interaction~\cite{Amar2005LowlevelCO}. In this question, participants are required to estimate the vaccinated portion of people in a target country.
  
  Hard question example: \textit{Approximately what portion of people are vaccinated in Bhutan?} Hard responses: \textit{less than $\frac{1}{3}$, between $\frac{1}{3} - \frac{2}{3}$, more than $\frac{2}{3}$}. 
  \vspace{-0.1in}
\end{description}

\subsubsection{Part 2 Measures}

Four possible chart pairings, times two possible layouts (e.g., Vertical or Radial chart on either right or left), times two levels of data density, times three question types/difficulty levels made for a total of $42$ multiple-choice questions. To assess performance, we recorded the selected answer and the response time (RT) of each question in all parts of the study. Stimuli were shown randomly to participants to minimize learning effects.

\subsubsection{Motivations and Difficulty Assessment}\label{sec:motiv}

While not part of our main hypotheses, we also recorded participants' perception of difficulty of both parts of the study. This was assessed on a 5-point Likert scale ranging from Easy to Difficult. These questions were asked to ascertain if perception of difficulty is linked to performance in both parts, and to aid in explainability of outcomes for anomalies or unexpected results. 

Further, research shows that cognitive skills (and specifically spatial skills~\cite{Atit2020ExaminingTR, FERGUSON20151}) and personal motivations are closely related and can together explain achievement in STEM subjects~\cite{Atit2020ExaminingTR, Carrera2018TeachingWA, Glynn2009ScienceMQ, Bryan2011MotivationAA}. Motivation is part of an individual's goal structures, and their belief about what is important, and determines whether they will engage in a given activity~\cite{Ames1992ClassroomsGS}. Research shows that motivation and feelings towards mathematics (or math anxiety) are interrelated~\cite{Atit2020ExaminingTR}. Additionally, performance on spatial tasks, and tasks involving mathematics, are influenced by self-efficacy, math anxiety, and intrinsic and extrinsic motivators~\cite{Glynn2009ScienceMQ, Ashcraft2001TheRA, FERGUSON20151}. To increase explainability and psychologically grounded methods in visualization research, we adapted the Science Motivation Questionnaire (SMQ) from \cite{Glynn2009ScienceMQ}, which is grounded in prominent theories of academic motivation and encompasses many of the motivational factors found to influence performance in STEM~\cite{Atit2020ExaminingTR}. We chose one question from each of the motivational factors measured by the SMQ: intrinsic, extrinsic, self-efficacy, self-determination, and math anxiety. Each question mirrored the SMQ with the word ``science'' replaced by ``data visualization'' and was rated by participants on a 5-point Likert scale as per the questionnaire (see SM).

\begin{figure}[tb]
 \centering 
 \includegraphics[width=\columnwidth]{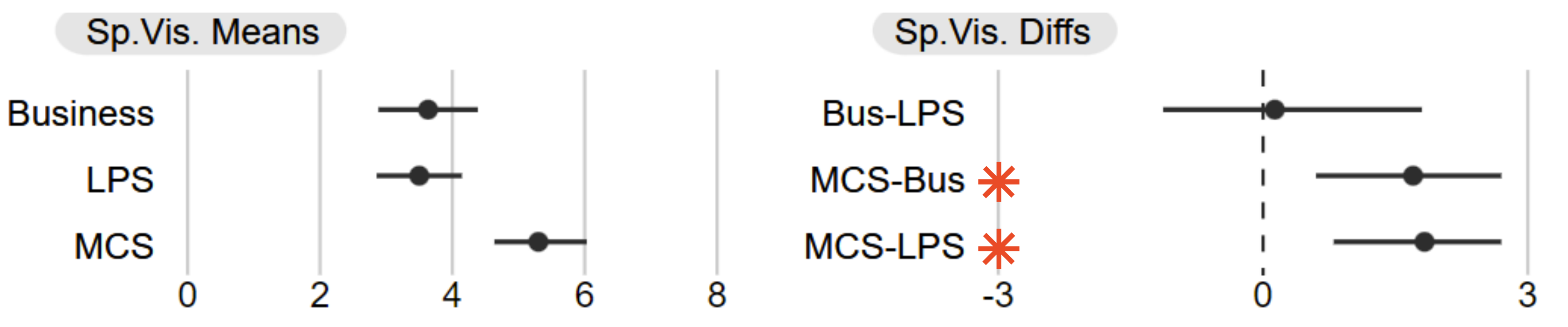}
 \caption{\textbf{H1}: Spatial visualization per domain, with the CI of means (left)
and of mean differences (right). Error bars represent 95\% Bootstrap confidence intervals. In the mean differences plot (right), those tighter and farther away from 0 provide stronger evidence of differences. Stars indicate evidence of significant differences across charts.}
 \label{fig:Results2}
\end{figure}

\begin{figure}[ht]
 \centering 
 \includegraphics[width=\columnwidth]{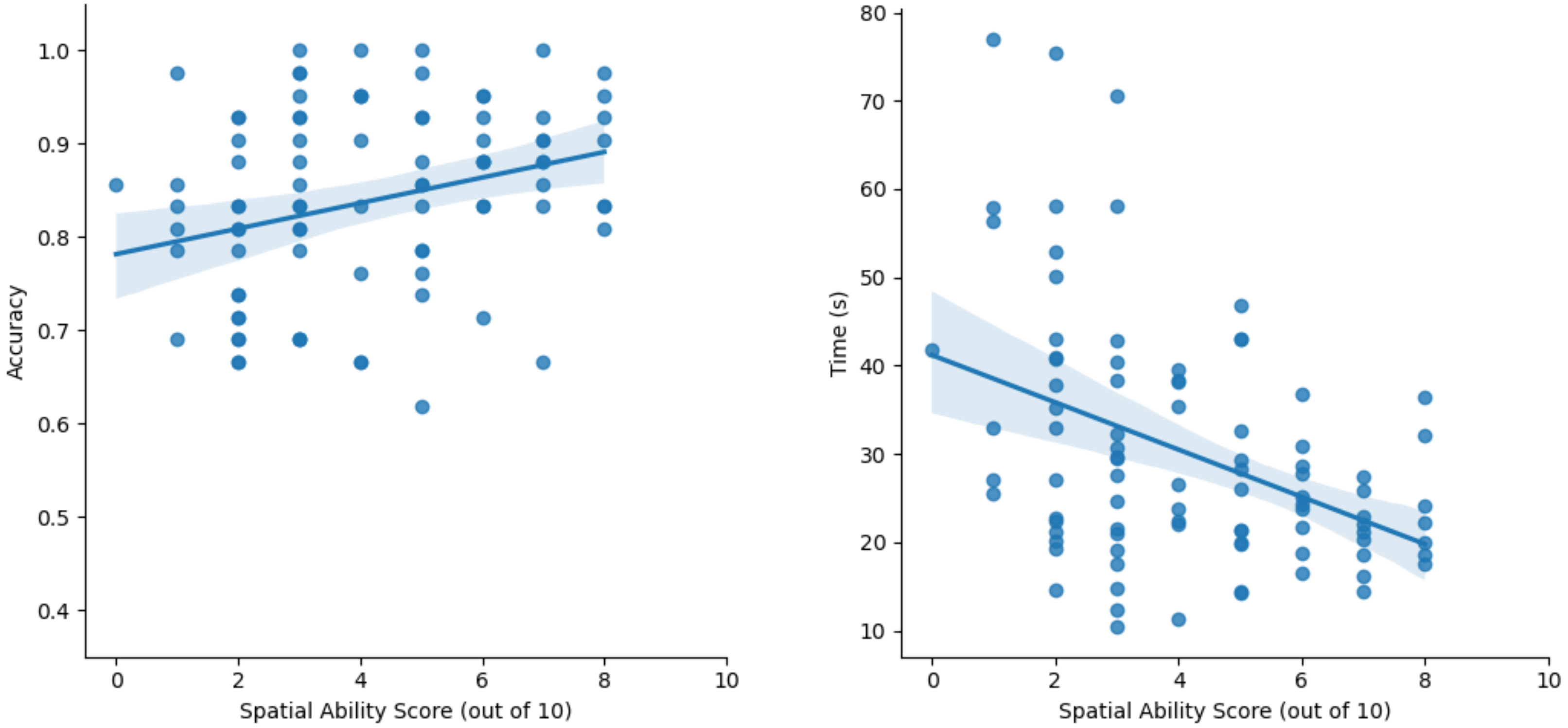}
 \caption{\textbf{H2}: Spatial visualization score with respect to performance: Pearson correlation between spatial visualization and accuracy (left) at R=.255, p=0.01. Pearson correlation between spatial visualization and response time (right) at R=-0.408, p=0.0006. Time in seconds.\vspace{-0.2in}}
 \label{fig:SpatialPearson}
\end{figure}




\subsection{Participants}
We successfully gathered 30 participants with reliable data in each target domain, giving us a balanced sample, for 90 total participants. We followed a strict rule of balanced samples across groups to support our between subject design.
The demographic statistics as a whole and across the three domains were as follows. Gender participation overall: 51\% male, 49\% female. Business: $47\%$ male, $53\%$ female. LPS: $30\%$ male, $70\%$ female. MCS: $77\%$ male, $23\%$ female. These are consistent with known educational domain gender differences across Europe~\cite{eurostat_2020,VidalFernandez}. The average age ($\pm$ standard deviation) of all participants was $24\pm4$, $23\pm4$ for Business, $25\pm5$ for LPS, and $23\pm4$ for MCS. In an effort toward trans-cultural research design in data visualization~\cite{CHIAbstract, GarciaDecol}, we opened our study to respondents from all countries. To this end we deployed our study across several days and different time slots to cater for time zone differences. Overall, we had $49\%$ from the Global South and $51\%$ from the Global North. Business:  $57\%$ Global South, $43\%$ Global North. LPS: $57\%$ Global South, $43\%$ Global North. MCS: $33\%$ Global South, $67\%$ Global North. 

\section{Results}

We analyzed differences in performance and spatial visualization using sample means, hypothesis testing at the $p < 0.05$ level of confidence, and 95\% confidence intervals. Confidence intervals (CIs) were constructed in Python using bias-corrected and accelerated bootstrapping (BCa) with 5000 iterations. After testing independent groups for normalcy, we decided to utilize BCa to create confidence intervals along with hypothesis testing using the Mann-Whitney U test to obtain a test statistic and p-value for further validation, as both are robust to non-parametric data~\cite{nachar2008mann, BootstrapJung, Puth2015OnTV}. We use both to demonstrate multiple analysis techniques and strength of evidence about the population means, as recommended in recent reviews~\cite{Greenland2016StatisticalTP, Dragicevic2016}. Confidence intervals allow for both traditional statistically significant interpretation (if the interval does not overlap 0), as well as subtle differences: the farther from 0 and the tighter it is, the stronger the evidence~\cite{Tan2010TheCI}. Additionally, we analyzed differences in spatial visualization using a Pearson's correlation test, where $p < 0.05$ indicates correlation and R ranging $0.1-0.3$ indicates weak, $0.3-0.5$ moderate, and $0.5-1.0$ strong correlation~\cite{Cohen1969StatisticalPA}. Below we report on our high-level findings; detailed means and stepwise analysis are reported in SM due to space restrictions.

\subsection{Demographic Differences}\label{sec:demogdiff}
While not part of our main hypotheses, we evaluated spatial visualization and accuracy levels across demographics.

Between self-identified males and females, we found no significant differences in spatial visualization level, keeping with previous research~\cite{Atit2020ExaminingTR, schooten_exploring_2009}. While we detected no significant difference in RT, there was a significant difference of $7\%$ in their accuracy (CI$(2,11)$, $p<0.01$). We note that the uneven distribution of men in MCS is likely a driving factor in this difference~\cite{jones_spatial_2008,tosto_why_2014}. 

We found strong evidence ($p<0.001$) that those who were raised in the Global South score lower on spatial visualization than those in the Global North (by $2/10$ questions). Additionally, overall mean times per question were 7.2s longer (CI$(4,9)$, $p<0.01$) and overall accuracy $5\%$ lower for those from the Global South ($p=0.05$). These differences reflect current research in psychology and information and communication technology demonstrating differing levels of cognitive ability and visualization interpretation between those in the Global South and North~\cite{Eppig2010ParasitePA, Figueredo2020TheBO, Rushton2005THIRTYYO,Anik2015SpatialDV}. 

\subsection{H1: Spatial Visualization by Discipline}
We hypothesized that spatial visualization ability would differ across discipline with MCS having the highest spatial visualization followed by Business, then LPS. 

Those in MCS had a mean spatial visualization score of $5.3$. There is compelling evidence that MCS had higher spatial visualization levels than Business ($p<0.01$) and LPS ($p<0.001$) by on average $2/10$ for both. Business had a mean score of $3.6$ while LPS had a mean score of $3.5$ -- there was no evidence of difference between them. The CI of mean differences between MCS and Business is $(0.6,2.7)$ while the CI for MCS and LPS is $(0.8,2.7)$, indicating a slightly stronger difference between MCS and LPS (see Fig.\ref{fig:Results2}). Though Business and LPS have slightly differing spatial abilities according to \cite{Wai2009SpatialAF}, our finding is in line with \cite{Shea2001ImportanceOA} which does not show a significant difference between spatial abilities of Business and LPS. Similar to both \cite{Wai2009SpatialAF, Shea2001ImportanceOA}, we found greater mean differences between MCS and Business/LPS than between Business and LPS themselves. See Section \ref{sec:motivresults} for further insight.

\smallskip
\noindent $\Rightarrow$ We partially confirmed \textbf{H1}, spatial visualization of participants differed for some domains. The ranking of disciplines in terms of spatial visualization is LPS $\approx$ Business $<$ MCS. This finding is still consistent with previous research~\cite{Shea2001ImportanceOA, Wai2009SpatialAF} where LPS and Business have similarly low levels of spatial ability, however we found no difference where they find that LPS has slightly lower levels than Business. 

\begin{figure*}[!ht]
  \begin{floatrow}
     \ffigbox{\includegraphics[width=.45\textwidth]{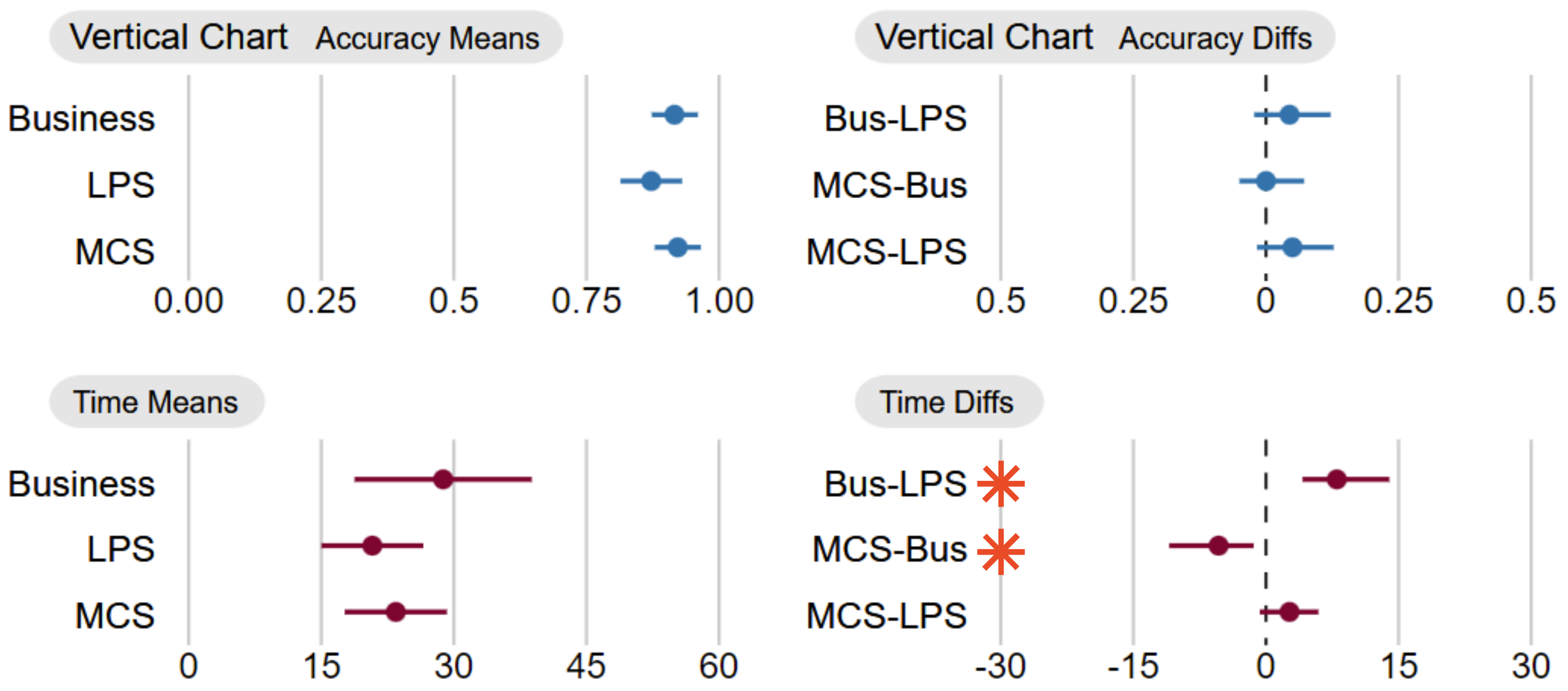}}{\caption{\textbf{H3}: Vertical Bars performance means (left) and mean differences (right), for Accuracy and Response Time. Time in seconds.}\label{fig:vert}}
     \ffigbox{\includegraphics[width=.45\textwidth]{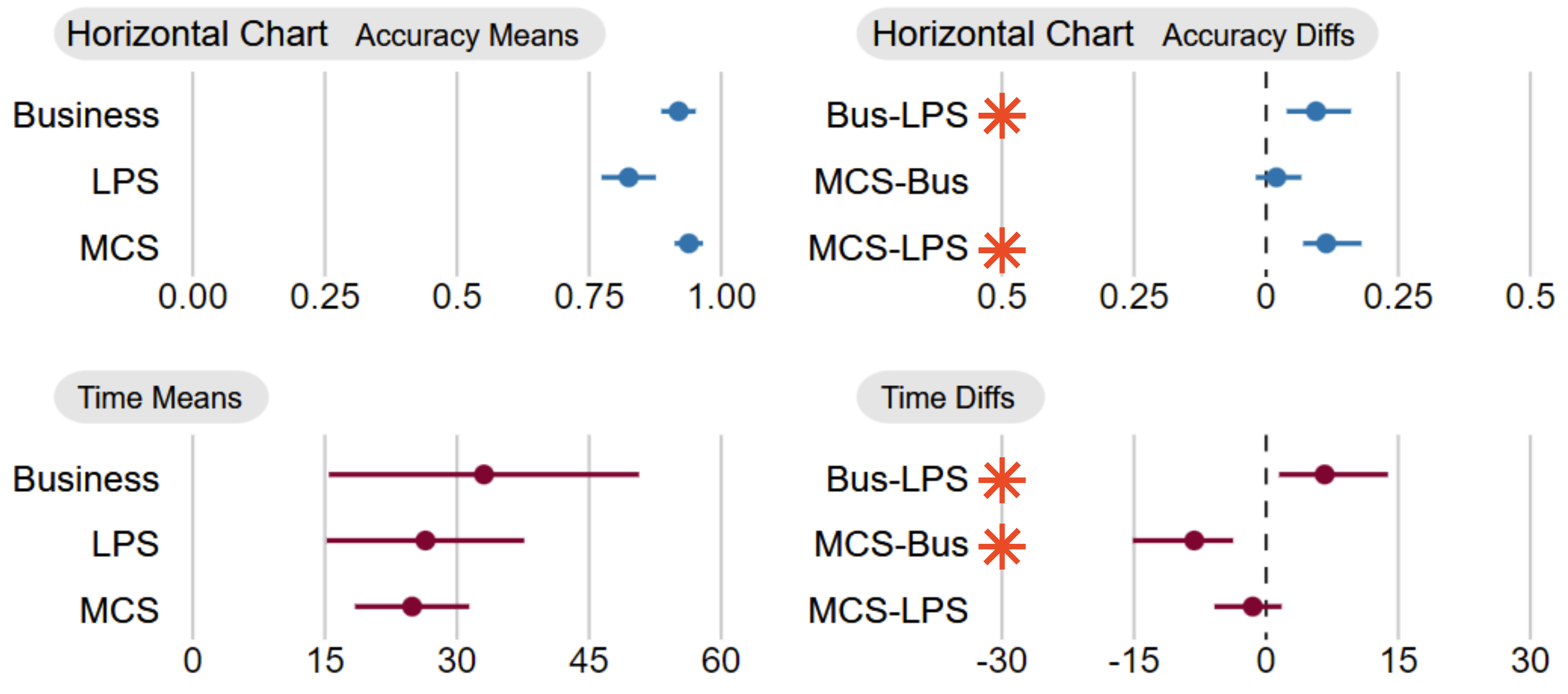}}{\caption{\textbf{H3}: Horizontal Bars performance means (left) and mean differences (right), for Accuracy and Response Time. Time in seconds.}\label{fig:horiz}}
  \end{floatrow}
\end{figure*}

\begin{figure*}[!ht]
  \begin{floatrow}
     \ffigbox{\includegraphics[width=.45\textwidth]{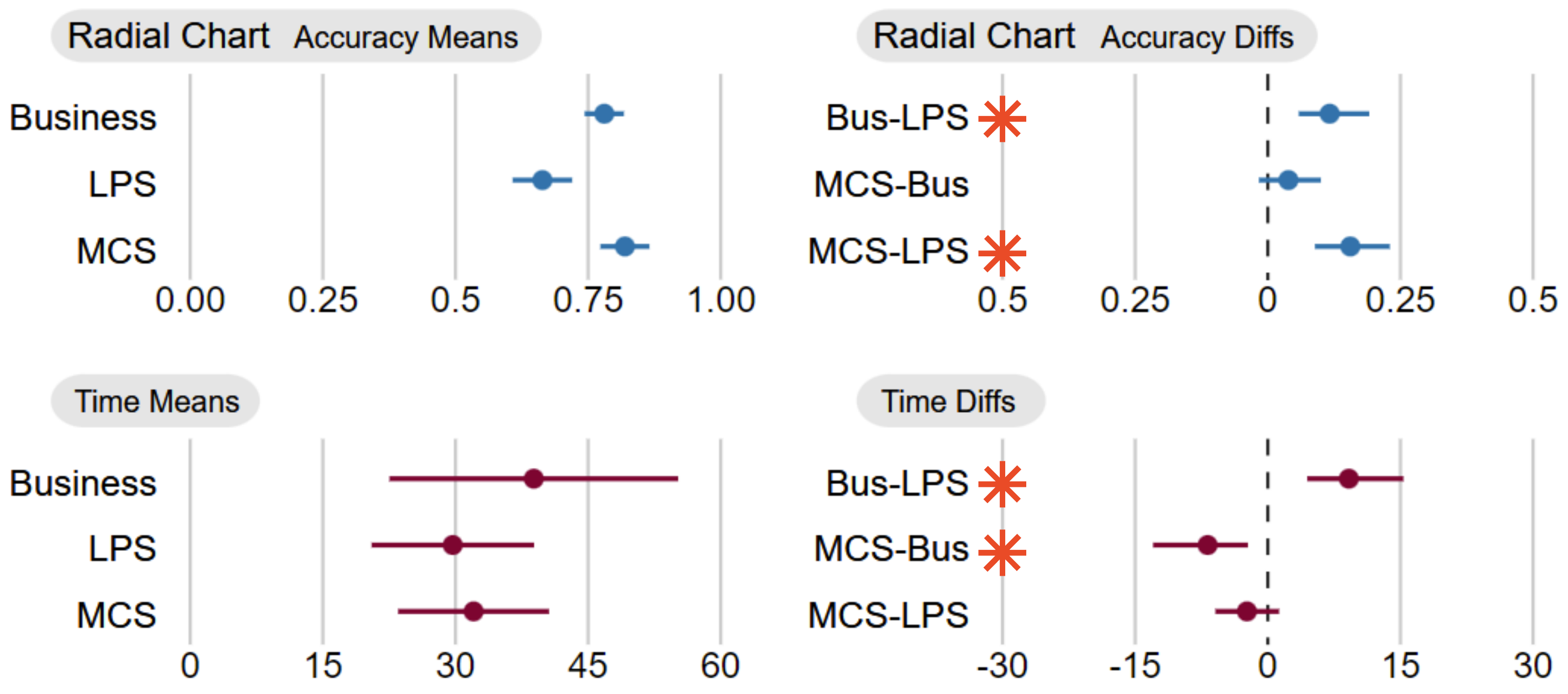}}{\caption{\textbf{H3}: Radial Bar Chart performance means (left) and mean differences (right), for Accuracy and Response Time. Time in seconds.\vspace{-0.3in}}\label{fig:rad}}
     \ffigbox{\includegraphics[width=.45\textwidth]{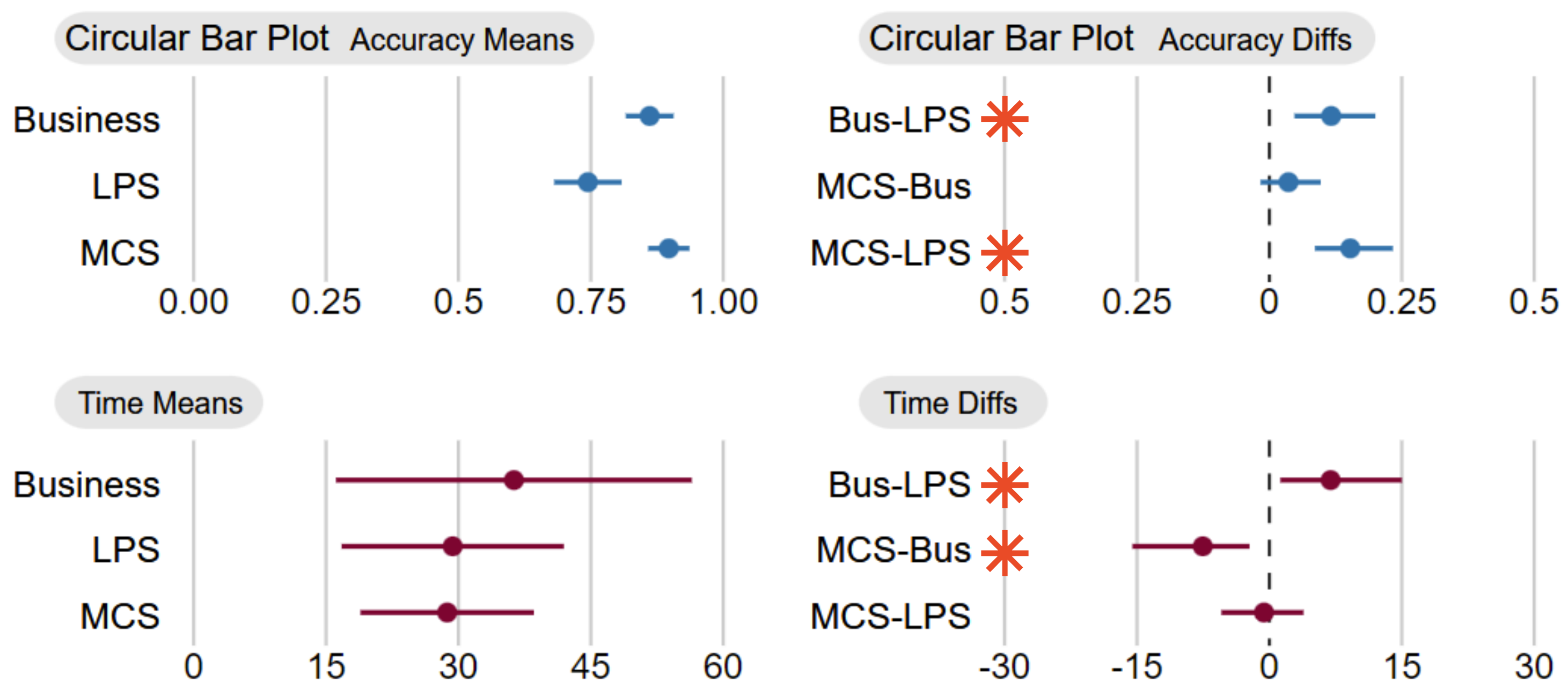}}{\caption{\textbf{H3}: Circular Bar Plot performance means (left) and mean differences (right), for Accuracy and Response Time. Time in seconds. \vspace{-0.3in}}\label{fig:star}}
  \end{floatrow}
\end{figure*}

\subsection{H2: Performance by Spatial Visualization}
We hypothesized that performance (accuracy and response time) would differ between those with high and low spatial visualization. To determine this, we look at both the correlation between spatial visualization and performance, as well as tested sample means and CIs across densities, charts, and questions. For all 90 participants, the mean spatial visualization score was $4/10$. We had 51 participants with low spatial visualization (9 MCS, 22 Business, 20 LPS) and 39 with high spatial visualization (21 MCS, 8 Business, 10 LPS). Low spatial individuals had a mean score of $2.5/10$, while High had $6/10$, meaning high spatial levels were about $4/10$ higher (CI$(3.3,4.1)$, $p<0.001$).  


We found evidence ($p\leq0.01$) of moderate correlation as spatial visualization increases with both accuracy (R=0.3) and RT (R=0.4) (see Fig. \ref{fig:SpatialPearson}). 
We compared performance of low and high spatial visualization participants across the two density levels of 7 and 14 data points, 4 chart types and 3 density levels. We report results from our analysis below (SM reports detailed breakdowns of sample means and CIs).


\subsubsection{By Density}
\textbf{Response Time:}
Mean times were higher for low spatial individuals compared to high spatial for both densities: 8.4s (CI$(6,12)$, $p<0.001$) higher for charts with 7 data points and 9.6s (CI$(7,13)$, $p<0.001$) higher for charts with 14 data points.

Within low spatial individuals, charts with 14 data points took 4.1s longer than charts with 7 data points (CI$(2,8)$, $p<0.001$). High spatial individuals took 2.9s longer on charts with 14 data points (CI$(1,5)$, $p<0.001$) compared to charts with 7 data points.

\textbf{Accuracy:}
We detected a slightly higher accuracy means on density 7 charts for high spatial individuals than low spatial by 5.2\% (CI$(0.7,10)$, $p<0.05$). No significant accuracy differences were detected between or within low and high spatial individuals across chart density. 

\subsubsection{By Chart Type}
\textbf{Response Time:}
For Vertical charts, low spatial individuals took 6.4s (CI$(3,10)$, $p<0.001$) longer than high spatial; for Horizontal low spatial took 9.7s (CI$(6,14)$, $p<0.001$) longer; for Radial they took 7.6s (CI$(4,12)$, $p<0.001$) longer; and for Circular Bar Plots, low spatial took 11.0s (CI$(7,16)$, $p<0.001$) longer.

Within low spatial individuals, polar coordinate charts took 6.0s longer than Cartesian coordinates (CI$(2,10)$, $p<0.001$). Within high spatial individuals, polar coordinate charts took 5.2s longer than Cartesian coordinates (CI$(4,7)$, $p<0.001$).

\textbf{Accuracy:}
There is some evidence that on Vertical charts low spatial means were 5.7\% (CI$(0.3,11)$, $p<0.05$) lower than high spatial. For Horizontal charts, low spatial means were 5.1\% (CI$(0.6,10)$, $p<0.05$) lower. On Circular Bar plots low spatial was 7.8\% (CI$(2,14)$, $p<0.05$) lower. There was no accuracy difference detected for Radial charts.

Within low spatial individuals, polar coordinate chart means were 10.1\% lower than Cartesian coordinate charts (CI$(5,15)$, $p<0.001$). Within high spatial individuals, polar coordinate chart means were similarly 10.5\% lower than Cartesian charts (CI$(6,15)$, $p<0.001$).


\subsubsection{By Question Difficulty}
\textbf{Response Time:}
There is compelling evidence that RT means were higher for low spatial individuals across all levels of difficulty: by 8.7s for Easy (CI$(6,12)$, $p<0.001$), by 7.5s for Medium (CI$(4,12)$, $p<0.01$), and by 10.7s for Hard (CI$(7,15)$, $p<0.001$). 

Within low spatial individuals, RT means were higher for Medium by 5.7s (CI$(2,10)$, $p<0.001$) and Hard by 7.0s (CI$(3,12)$, $p<0.001$) when compared to Easy time means, with no significant difference between Medium and Hard questions. For high spatial individuals, time means were also higher for Medium by 7.0s (CI$(5,9)$, $p<0.001$) and Hard by 5.1s (CI$(3,7)$, $p<0.001$) when compared to Easy times. Interestingly, high spatial time means for Hard questions were 1.9s \textit{lower} than for Medium times (CI$(0.5,4.3)$, $p<0.01$). 

\textbf{Accuracy:}
Accuracy was similar between low and high spatial individuals in Easy and Medium questions. However, the low spatial means were lower by 10.1\% (CI$(4,17)$, $p=0.01$) for Hard questions. 

Within low spatial individuals, accuracy means were 11.9\% lower for Medium (CI$(6,17)$, $p<0.001$) and 17.8\% lower for Hard (CI$(11,24)$, $p<0.001$) when compared to Easy questions. High spatial accuracy means were 16.1\% lower for Medium (CI$(11,22)$, $p<0.001$) and 12.3\% lower for Hard (CI$(9,17)$, $p<0.001$) when compared to Easy questions. Neither had notable accuracy differences between Medium and Hard questions within group. 

\smallskip
\noindent $\Rightarrow$ When looking at overall performance, we confirmed \textbf{H2}, spatial visualization level does affect RTs (lower spatial visualization, longer time to answer questions) and a moderate amount of accuracy. Low and high spatial individuals have a similar discrepancy in performance as density increases and coordinates move between Cartesian and polar.

\subsection{H3: Performance by Domain}
We hypothesized that performance (accuracy and response time) would differ between disciplines, with MCS performing best and LPS worst (following their spatial ability). Looking at overall mean differences, we found that MCS was faster than Business by 7.2s (CI$(5,10)$, $p<0.001$), and LPS was faster than Business by 7.6s (CI$(5,11)$, $p<0.001$). No difference was detected in overall RT between MCS and LPS. For overall accuracy however, MCS was 12.8\% more accurate than LPS (CI$(9,19)$, $p<0.001$), and Business was 10\% more accurate than LPS (CI$(5,16)$, $p<0.001$). No difference was detected in overall accuracy between MCS and Business. 

We detail results of our analysis of accuracy and RT across three variables: density, chart type (Figs. \ref{fig:vert}-\ref{fig:star}), and question difficulty. Detailed breakdown of the analysis and means are provided in the SM.

\subsubsection{By Density}

\textbf{Response Time:}
For charts of density 7, there is evidence that mean times were higher for Business compared to LPS by 7.6s (CI$(4,13)$, $p<0.001$) and MCS by 7.3s (CI$(4,13)$, $p<0.001$). There is also evidence mean times were higher for Business in charts of density 14: higher than LPS by 7.7s (CI$(4,13)$, $p<0.001$) and MCS by 7.0s (CI$(3,12)$, $p<0.001$). There is no evidence of significant differences between MCS and LPS mean times. 

Within groups there was similar increased time means for charts of density 14 when compared to density 7: Business by 3.5s, LPS by 3.4s, MCS by 3.8s (all $p<0.001$). 

\textbf{Accuracy:}
For charts of density 7, there is strong evidence that MCS had 11.6\% higher accuracy (CI$(7,18)$, $p<0.001$) and Business had 9.8\% higher accuracy means (CI$(5,16)$, $p<0.001$) than LPS. Additionally, for charts of density 14, MCS had 14.0\% (CI$(9,20)$, $p<0.001$) higher accuracy and Business had 10.2\% (CI$(5,17)$, $p<0.001$) higher accuracy means than LPS. There is no evidence of differences between MCS and Business in accuracy. 

There is no evidence of accuracy differences between densities within groups.

\subsubsection{By Chart Type}

See Figs. \ref{fig:vert}, \ref{fig:horiz}, \ref{fig:rad}, and \ref{fig:star} for accuracy and RT means and difference CIs.

\textbf{Response Time:}
There is evidence that mean times were 8.0s higher for Business than LPS ($p<0.001$) and 5.4s higher than MCS ($p<0.05$) for Vertical charts. For Horizontal charts, Business took 6.6s longer than LPS ($p<0.001$) and 8.2s longer than MCS ($p<0.001$). Looking at Radial charts, Business took 9.2s longer than LPS ($p<0.001$) and 6.8s longer than MCS ($p<0.05$). Last, for Circular Bar Plots, Business took 6.9s longer than LPS ($p<0.001$) and 7.6s longer than MCS ($p=0.01$).

Within all domains, there was a similar increase in time between Cartesian and polar coordinate charts: Business by 5.9s (CI$(0.5,11)$, $p<0.001$), LPS by 5.0s (CI$(2,8)$, $p<0.001$), and MCS by 6.0s (CI$(4,8)$, $p<0.001$). 

\textbf{Accuracy:}
For Vertical charts, there is no evidence of accuracy differences. For Horizontal charts however, there is strong evidence that mean accuracy of MCS was 11.4\% higher than LPS and Business was 9.4\% higher (both $p<0.001$). For Radial charts, there is evidence MCS and Business had higher accuracy means than LPS (by 15.6\% and 11.7\%, respectively, at $p<0.001$). For Circular Bars MCS means were higher by 15.3\% and Business means higher by 11.7\% than LPS, again at $p<0.001$. There is no significant evidence of difference in means between MCS and Business across chart types.  

Within all domains, accuracy means were similarly lower for polar coordinate plots when compared to Cartesian coordinate charts: Business by 9.8\% (CI$(5,15)$, $p<0.001$), LPS by 13.7\% (CI$(6,20)$, $p<0.001$), and MCS by 7.5\% (CI$(3,12)$, $p<0.01$).  

\subsubsection{By Question Difficulty}
\textbf{Response Time:}
For Easy questions, there is evidence of increased mean times for Business compared to LPS by 5.4s (CI$(2,10)$, $p=0.001$) and MCS by 7.2s (CI$(4,12)$, $p<0.01$). Business also has increased times for Medium questions compared to LPS by 9.3s (CI$(5,17)$, $p<0.001$) and MCS by 7.3s (CI$(3,15)$, $p=0.01$). For Hard questions, Business had increased times by 8.2s compared to LPS (CI$(3,14)$, $p<0.001$) and 7.0s to MCS (CI$(3,13)$, $p<0.01$). There is no evidence of differences in time means for LPS and MCS.    

Within Business there were increased times of 7.6s to Medium questions (CI$(2,15)$, $p<0.001$) and 7.0s to Hard questions (CI$(2,13)$, $p<0.001$) from Easy questions. For LPS there were increased times of 3.7s to Medium questions (CI$(0.6,7)$, $p=0.01$) and 4.3s to Hard questions (CI$(0.5,9)$, $p<0.05$) from Easy questions. For MCS there were increased times of 7.6s to Medium questions (CI$(5,11)$, $p<0.001$) and 7.2s to Hard questions (CI$(5,10)$, $p<0.001$) from Easy questions. No differences were detected in mean times between Medium and Hard questions across domains.

\textbf{Accuracy:}
For Easy questions there is some evidence that MCS had higher accuracy than LPS by 9.8\% (CI$(5,18)$, $p<0.01$) and Business by 3.1\% (CI$(0.5,6)$, $p<0.05$). Additionally, there was no detected difference between Business and LPS accuracy means for Easy nor Medium questions. There is some evidence that MCS had 11.9\% higher scores than LPS in Medium questions (CI$(4,20)$, $p<0.01$). For Hard questions, Business had higher accuracy means than LPS by 16.4\% (CI$(8,25)$, $p=0.01$) and MCS was higher than LPS by 16.7\% (CI$(8,25)$, $p=0.01$). Other than in Easy questions there was no difference detected between MCS and Business. 

Within Business, accuracy means fell by 14.2\% (CI$(10,20)$, $p<0.001$) from Easy to Medium questions and by 11.2\% (CI$(6,17)$, $p<0.001$) from Easy to Hard. Within LPS accuracy means fell by 14.5\% (CI$(5,23)$, $p<0.001$) from Easy to Medium questions and by 20.1\% (CI$(11,30)$, $p<0.001$) from Easy to Hard. For MCS accuracy means fell by 12.4\% (CI$(7,18)$, $p<0.001$) from Easy to Medium questions and by 14.0\% (CI$(10,19)$, $p<0.001$) from Easy to Hard. There was no evidence of differences between Medium and Hard questions within domains.

\smallskip
$\Rightarrow$ Our results partially confirm \textbf{H3}. MCS participants tend to have higher accuracy than LPS and lower RTs than Business participants. MCS has higher accuracy than LPS across density, chart types (save Vertical charts), and question difficulty; MCS also has faster times than Business across density, chart types, and question difficulty.  However, although spatial visualization levels are similar for Business and LPS, Business participants tend to have higher RTs and higher accuracy than LPS. Business and LPS did not differ significantly in accuracy only on Easy \& Medium questions, and Vertical charts -- although LPS was faster. Overall, we observed a time/error trade-off between Business and LPS -- the more time spent, the higher the accuracy. This finding is in line with~\cite{Hall2022ProfessionalDA} that found time/error trade offs between domains with similar levels of spatial visualization (CS and Education). However, in our study no trade off was detected for MCS participants (who had higher spatial visualization levels) which were generally just as fast as LPS and just as accurate as Business. The difference in results could potentially be linked to the fact that we compared across a balanced sample of participants and included education background to professional profile in case profession was misaligned, where \cite{Hall2022ProfessionalDA} did not. This indicates cognitive abilities do interact with domain to affect performance in both accuracy and RT, and that accuracy can be improved with increased RTs. The latter finding is common in visualization research on spatial visualization \cite{toker_gaze_2019, Downing2005TheEA, Hall2022ProfessionalDA}.

\subsection{Difficulty and Motivation}\label{sec:motivresults}

\begin{table}[t]
\resizebox{\columnwidth}{!}{%
\begin{tabular}{|c|l|l|}
\hline
\textbf{Motivation} &
  \multicolumn{1}{c|}{\textbf{Mean Scores}} &
  \multicolumn{1}{c|}{\textbf{Significant Differences}} \\ \hline
\multirow{2}{*}{\begin{tabular}[c]{@{}c@{}}Overall \\ (out of 20)\end{tabular}} &
  \begin{tabular}[c]{@{}l@{}}High Spatial - 11.1 \rule{0pt}{2.3ex}   \\ Low Spatial - 11.7 \\ \end{tabular} &
  None \\ \cline{2-3} 
 &
  \begin{tabular}[c]{@{}l@{}}MCS - 12 \rule{0pt}{2.3ex}\\ Business - 12.3\\ LPS - 9.9\end{tabular} &
  \begin{tabular}[c]{@{}l@{}}MCS\textgreater{}LPS: CI$(0.6,4)$, $p<0.01$\\ Business\textgreater{}LPS: CI$(0.8,4)$, $p=0.01$\end{tabular} \\ \hline
\multirow{2}{*}{\begin{tabular}[c]{@{}c@{}}Intrinsic\\ (out of 4)  \\ {\small Spend my own time learning} \\ {\small about data visualization}\end{tabular}} &
  \begin{tabular}[c]{@{}l@{}}High Spatial - 1.4 \rule{0pt}{2.3ex}\\ Low Spatial - 2.0\end{tabular} &
  CI$(0.2, 1)$, $p<0.01$ \\ \cline{2-3} 
 &
  \begin{tabular}[c]{@{}l@{}}MCS - 1.8 \rule{0pt}{2.3ex}\\ Business - 1.9\\ LPS - 1.7\end{tabular} &
  None \\ \hline
\multirow{2}{*}{\begin{tabular}[c]{@{}c@{}}Extrinsic\\ (out of 4) \\ {\small My career or studies} \\ {\small involve data visualization}\end{tabular}} &
  \begin{tabular}[c]{@{}l@{}}High Spatial - 2.3 \rule{0pt}{2.3ex}\\ Low Spatial - 2.1\end{tabular} &
  None \\ \cline{2-3} 
 &
  \begin{tabular}[c]{@{}l@{}}MCS - 2.6 \rule{0pt}{2.3ex}\\ Business - 2.4\\ LPS - 1.6\end{tabular} &
  \begin{tabular}[c]{@{}l@{}}MCS\textgreater{}LPS: CI$(0.6, 2)$, $p<0.001$\\ Business\textgreater{}LPS: CI$(0.3,1)$, $p<0.01$\end{tabular} \\ \hline
\multirow{2}{*}{\begin{tabular}[c]{@{}c@{}}Self-Determination\\ (out of 4)\\ {\small I put effort into learning} \\ {\small about data visualization}\end{tabular}} &
  \begin{tabular}[c]{@{}l@{}}High Spatial - 2.2 \rule{0pt}{2.3ex}\\ Low Spatial - 2.9\end{tabular} &
  CI$(0.19,1.1)$, $p<0.01$ \\ \cline{2-3} 
 &
  \begin{tabular}[c]{@{}l@{}}MCS - 2.2 \rule{0pt}{2.3ex}\\ Business - 3.0\\ LPS - 2.5\end{tabular} &
  \begin{tabular}[c]{@{}l@{}}Business\textgreater{MCS}: CI$(0.3,1)$, $p<0.01$\\ Business\textgreater{}LPS: CI$(0.03,1)$, $p<0.05$\end{tabular} \\ \hline
\multirow{2}{*}{\begin{tabular}[c]{@{}c@{}}Self-Efficacy\\ (out of 4)\\ {\small I am confident I will perform} \\ {\small well on data visualization tasks}\end{tabular}} &
  \begin{tabular}[c]{@{}l@{}}High Spatial - 2.7 \rule{0pt}{2.3ex}\\ Low Spatial - 2.7\end{tabular} &
  None \\ \cline{2-3} 
 &
  \begin{tabular}[c]{@{}l@{}}MCS - 2.8 \rule{0pt}{2.3ex}\\ Business - 2.9\\ LPS - 2.5\end{tabular} &
  \begin{tabular}[c]{@{}l@{}} Business\textgreater{}LPS: CI$(0.03,0.8)$, $p<0.05$\end{tabular} \\ \hline
\multirow{2}{*}{\begin{tabular}[c]{@{}c@{}}Low Math Anxiety \\ (out of 4) \\ {\small Note: higher score} \\ {\small indicates lower anxiety} \end{tabular}} & 
  \begin{tabular}[c]{@{}l@{}}High Spatial - 2.4 \rule{0pt}{2.3ex}\\ Low Spatial - 1.9\end{tabular} &
  None \\ \cline{2-3} 
 &
  \begin{tabular}[c]{@{}l@{}}MCS - 2.6 \rule{0pt}{2.3ex}\\ Business - 2.1\\ LPS - 1.7\end{tabular} &
  MCS\textgreater{}LPS: CI$(0.2,2)$, $p=0.01$ \\ \hline
\end{tabular}%
}
\caption{Motivation scores and significant differences between spatial visualization level and education domain. Details in SM.\vspace{-0.2in}}
\label{tab:mot-table}
\end{table}

The final tasks of the study were to rate the difficulty of each section along with rating personal motivations regarding data visualization. 
High spatial individuals rated the spatial visualization assessment (out of 4) $1.7/4$ on average, while low spatial individuals rated it $2.19/4$ - this was significantly different at $p<0.05$ with CI$(0,0.94)$. No differences were detected between the low and high spatial ratings of Part 2. The only significant difference found between domains was the average difficulty rating of the spatial visualization assessment between LPS ($2.5/4$) and MCS ($1.7/4$) at $p<0.05$ with CI$(0,1.1)$. Business rated the spatial visualization assessment $(1.93/4)$ on average but was not significantly different from LPS nor MCS. There were no differences detected between the domain's average rating of Part 2. 

As per the SMQ, each motivation response corresponded to a score from 1-4, rating agreement with the statement~\cite{Glynn2009ScienceMQ}. The statement measuring anxiety was reverse scored so that higher scores correspond to lower anxiety - thus, we refer to this construct as low math anxiety. See Table \ref{tab:mot-table} for motivation ratings and significant differences. 

The overall average rating of data visualization motivation was $11.4/20$. Those above the average motivation score had quicker RTs overall by 7.1s than those below the average (CI$(5,10)$, $p<0.001$). They were significantly faster across densities, chart types, and question types (all $p<0.001$). Additionally, those above the average had higher accuracy in charts of density 14 ($p<0.05$) and polar coordinate charts ($p<0.05$) - see SM for details. 

As seen in Table \ref{tab:mot-table}, low spatial individuals have higher intrinsic motivation and self-determination than high spatial individuals. This finding is in line with research demonstrating that those with low visual processing cognitive abilities still choose visual data interfaces i.e., preference to use visualizations is not directly indicative of visual cognitive abilities~\cite{lee_correlation_2019}. 

Interestingly, both MCS and Business had significantly higher overall motivation scores compared to LPS and had higher accuracy overall. Specifically, Business outranked LPS in extrinsic motivation, self-efficacy, and self-determination (where Business also outranked MCS). MCS outranked LPS only in extrinsic motivation and low math anxiety. This indicates data visualization is more prevalent in MCS and Business domains than LPS and thus there may be a higher value placed on accuracy in visualization tasks. Additionally, putting effort into learning about data visualization and confidence in tasks may increase accuracy -- this is aligned with \cite{GANLEY2011235, collins1985self} that demonstrate that higher self-determination, self-efficacy, and confidence are correlated with increased performance in mathematics tasks.

\section{Discussion}


Our results confirm and build upon research in information visualization that both spatial visualization and domain influence use of data visualization~\cite{vanderplas_spatial_2016, KirbyCollab}. These individual differences are often treated independently, but we build on work that brings them together to increase effectiveness and reach of visualization~\cite{Hall2022ProfessionalDA}. We ensured balanced samples between domains, aimed at gender and global diversity, and included education background in the professional profile to reduce potential confounding factors. We found that there are differences between domains in statistical performance and motivations around data visualization, leading to implications and insights for design. 

\paragraph{Spatial visualization performance} We investigated how spatial visualization ability and discipline come together to affect task performance (accuracy and RT) on common data visualizations. We confirmed past research that spatial abilities vary with domain~\cite{Wai2009SpatialAF}, with MCS having significantly higher spatial visualization than Business and LPS (Fig. \ref{fig:Results2}, \textbf{H1}). 

Consistent with visualization research into spatial visualization \cite{vanderplas_spatial_2016,wenhong_user_2019,vicente_assaying_1987}, we found increased level of spatial visualization correlates with higher accuracy and quicker RTs (\textbf{H2}). This finding speaks to space in information visualization for inclusivity, given learning disorders and populations that have been connected to low spatial abilities~\cite{broitman_nvld_2020,Figueredo2020TheBO}. RTs differed amongst all sub-conditions (density, chart type, and question difficulty) and accuracy differed for Hard questions amongst all chart types between spatial visualization levels. We found a similar discrepancy in performance between Cartesian and polar coordinate charts within low and high spatial individuals. Additionally, we found that for Hard questions (involving math computation) high spatial individuals took less time than with Medium questions, but they maintained significantly higher accuracy than low spatial. This supports research in education psychology that high spatial individuals tend to perform better in mathematics~\cite{Young2018TheCB, Atit2020ExaminingTR}.

\textit{Difficulty \& Motivation:} Low spatial individuals found the spatial visualization assessment more difficult than high spatial, did not perceive the multiple-choice questions as more difficult, yet did not perform as well. Additionally, Low spatial individuals had higher levels of intrinsic motivation and self-determination (see Table \ref{tab:mot-table}). Both findings support \cite{wenhong_user_2019}, that individuals are often unaware of their abilities and still choose to interact with and enjoy visual data representations.

\paragraph{Domain Performance} 

Between domains, we found behavior consistent with spatial visualization level and performance (\textbf{H3}) such that MCS had higher accuracy than LPS and quicker RTs than Business. This difference was consistent across sub-conditions. This finding is consistent with research into psychology and visualization that individuals who studied STEM subjects have increased spatial abilities and performance in visualization tasks~\cite{Atit2020ExaminingTR, vanderplas_spatial_2016}. 

\textit{Difficulty \& Motivation:} Consistent with spatial visualization levels, LPS found the spatial visualization assessment more difficult than MCS, with Business rating difficulty between the two. We found an overall time/error trade-off with Business and LPS (i.e., the more time spent on questions the higher the accuracy). We conjecture this trade-off, and difficulty perception, has to do with the motivations and domain differences between Business and LPS. Literature suggests there is higher value placed on data visualization in Business~\cite{few2007data,zheng2017data, tegarden1999business, luo2019user} when compared to LPS~\cite{zinovyev2010data, henshaw2018data}. This is reflected in the motivations of Business and LPS participants in our study (see Table \ref{tab:mot-table}). Notably, MCS and Business reported that data visualization is involved in their career or studies to a higher degree than LPS. Business additionally rated confidence in doing well on visualization tasks as higher than LPS and reported they put more effort into learning about data visualization than both MCS and LPS. These motivations have been shown to correlate with increased performance in mathematics and engagement, performance, and higher quality learning in education in general~\cite{Atit2020ExaminingTR, Ryan2000SelfdeterminationTA}. From this, it follows that even with similar levels of spatial visualization, those in the Business domain would take time to have higher accuracy on data visualization tasks where those in LPS might not.

The fact that LPS exhibit different abilities and performance should be significant to the visualization community. Individuals in LPS are regularly at the center of government systems and are often advised by scientific bodies and advisors, especially during the ongoing COVID-19 pandemic~\cite{Colmane006928}. There could be broad reaching consequences if designers fail to cater to the abilities of those making high impact governance decisions when sharing scientific information. Given bodies like the World Health Organization, governments, universities, and data visualization tech companies created COVID-19 dashboards with spatially rotated data~\cite{harvardCOVID-19,msoftcovid-19,tabcovid-19,ibmcovid,whocovid}, it is critical that information visualization is accessible and effective for an imperative group of individuals with huge influence on society.

\paragraph{Outlook \& Future Work}
The scope of this study did not include the causal origins of the different spatial visualization abilities amongst disciplines - this is still an open question. It is important to note that other factors can influence task performance such as domain knowledge, representational fluency, visual familiarity, or demographic differences~\cite{KirbyCollab,Hall2022ProfessionalDA,Figueredo2020TheBO}; the interplay between these factors is complex and none alone can explain differences. Regardless of the origins of the differences, our study demonstrates how spatial visualization level and domain affect visualization use, which leads to important design implications. Further studies can be done on additional domains to expand knowledge of performance differences in visual tasks and spatial visualization between groups for higher impact designs. To increase inclusive design practices, interventions might be studied that allow low spatial individuals to increase accuracy and/or decrease RT needed for visual tasks: benefiting those in LPS and Business domains. Our finding that spatial visualization level is an important cause behind difference in performance is reflected in intervention research for cognitive abilities~\cite{conati_eye_2014, Caren_highlighting}. Performance including interventions and guided interactions (e.g., explicit linking across charts, feature highlighting, information redundancy) could be tested and balanced with cognitive load drawbacks for effectiveness. 

Our study advances initial work in the visualization community toward cataloguing cognitive differences of domains~\cite{Hall2022ProfessionalDA} and is a step toward increasing impact of visual design across disciplines.

\section{Conclusion}
The aim of our research was to build on work in visualization exploring spatial visualization differences amongst domains and its effect on visualization use. Our study expanded research to the domains of Business and LPS and included real-world data and visual scenarios of spatially rotated data. We presented additional evidence that the interplay between cognitive and demographic factors should be considered to increase effectiveness and inclusivity of visual design. Additionally, our research showed motivational differences between domains that could affect interaction and design needs. As the field of visualization evaluates cognitive and domain differences and the interplay between the two, studies like this can ensure informed and effectual design for all communities.


\bibliographystyle{abbrv-doi}
\newcommand{\tvcg}{IEEE Trans. Vis. \& Comp. Grap.}
\newcommand{\acmchi}{{ACM} Conf. Human Factors - {CHI}}
\newcommand{\cga}{IEEE Computer Graphics \& Applications}
\bibliography{template.bbl}

\begin{thebibliography}{10}

\bibitem{Ainooson2017ACM}
J.~Ainooson and M.~Kunda.
\newblock A computational model for reasoning about the paper folding task
  using visual mental images.
\newblock {\em Cognitive Science}, 2017.

\bibitem{CHIAbstract}
A.~Alvarado~Garcia, J.~F. Maestre, M.~Barcham, M.~Iriarte, M.~Wong-Villacres,
  O.~A. Lemus, P.~Dudani, P.~Reynolds-Cu\'{e}llar, R.~Wang, and
  T.~Cerratto~Pargman.
\newblock Decolonial pathways: Our manifesto for a decolonizing agenda in hci
  research and design.
\newblock In {\em \acmchi}. New York, NY, USA, 2021.

\bibitem{GarciaDecol}
A.~Alvarado~Garcia, J.~F. Maestre, M.~Barcham, M.~Iriarte, M.~Wong-Villacres,
  O.~A. Lemus, P.~Dudani, P.~Reynolds-Cu\'{e}llar, R.~Wang, and
  T.~Cerratto~Pargman.
\newblock Decolonial pathways: Our manifesto for a decolonizing agenda in hci
  research and design.
\newblock In {\em \acmchi}. Association for Computing Machinery, New York, NY,
  USA, 2021.

\bibitem{Amar2005LowlevelCO}
R.~A. Amar, J.~R. Eagan, and J.~T. Stasko.
\newblock Low-level components of analytic activity in information
  visualization.
\newblock {\em IEEE Symposium on Information Visualization, 2005. INFOVIS
  2005.}, pp. 111--117, 2005.

\bibitem{Ames1992ClassroomsGS}
C.~A. Ames.
\newblock Classrooms: Goals, structures, and student motivation.
\newblock {\em Journal of Educational Psychology}, 84:261--271, 1992.

\bibitem{Anik2015SpatialDV}
T.~A. Anik, K.~P. Dhali, S.~M.~F. Hossain, and F.~T. Johara.
\newblock Spatial data visualization methodologies in ict4d research.
\newblock {\em 2015 18th International Conference on Computer and Information
  Technology (ICCIT)}, pp. 429--434, 2015.

\bibitem{Ashcraft2001TheRA}
M.~H. Ashcraft and E.~P. Kirk.
\newblock The relationships among working memory, math anxiety, and
  performance.
\newblock {\em Journal of experimental psychology. General}, 130 2:224--37,
  2001.

\bibitem{ibmcovid}
S.~Assefa.
\newblock Gauteng province launches covid-19 dashboard developed by ibm
  research, wits university and gcro - now open to the public, 2020.

\bibitem{Atit2020ExaminingTR}
K.~Atit, J.~Power, N.~Veurink, D.~H. Uttal, S.~A. Sorby, G.~C. Panther,
  C.~Msall, L.~Fiorella, and M.~M. Carr.
\newblock Examining the role of spatial skills and mathematics motivation on
  middle school mathematics achievement.
\newblock {\em International Journal of STEM Education}, 7:1--13, 2020.

\bibitem{barral_understanding_2020}
O.~Barral, S.~Lall\'{e}, and C.~Conati.
\newblock Understanding the effectiveness of adaptive guidance for narrative
  visualization: A gaze-based analysis.
\newblock In {\em Proc. 25th Int. Conf. Intelligent User Interfaces (IUI)}, p.
  1–9, 2020.

\bibitem{borgo_empirical_2012}
R.~Borgo, A.~Abdul-Rahman, F.~Mohamed, P.~Grant, I.~Reppa, L.~Floridi, and
  M.~Chen.
\newblock An {Empirical} {Study} on {Using} {Visual} {Embellishments} in
  {Visualization}.
\newblock {\em \tvcg}, 18(12):2759--2768, Dec. 2012.

\bibitem{borgo_information_2018}
R.~Borgo, L.~Micallef, B.~Bach, F.~McGee, and B.~Lee.
\newblock Information {Visualization} {Evaluation} {Using} {Crowdsourcing}.
\newblock {\em Computer Graphics Forum}, 37(3):573--595, July 2018.

\bibitem{bostock_2016}
M.~Bostock.
\newblock d3.scaleradial · issue 90 · d3/d3-scale, 2016.

\bibitem{broitman_nvld_2020}
J.~Broitman, M.~Melcher, A.~Margolis, and J.~M. Davis.
\newblock {\em {NVLD} and {Devlopmental} {Visual}-{Spatial} {Disorder} in
  {Children}}.
\newblock Springer, 2020.

\bibitem{Bryan2011MotivationAA}
R.~R. Bryan, S.~M. Glynn, and J.~M. Kittleson.
\newblock Motivation, achievement, and advanced placement intent of high school
  students learning science.
\newblock {\em Science Education}, 95:1049--1065, 2011.

\bibitem{burch2014benefits}
M.~Burch and D.~Weiskopf.
\newblock On the benefits and drawbacks of radial diagrams.
\newblock In {\em Handbook of human centric visualization}, pp. 429--451.
  Springer, 2014.

\bibitem{Burnett1980EffectsOA}
S.~A. Burnett and D.~M. Lane.
\newblock Effects of academic instruction on spatial visualization.
\newblock {\em Intelligence}, 4:233--242, 1980.

\bibitem{Caren_highlighting}
G.~Carenini, C.~Conati, E.~Hoque, B.~Steichen, D.~Toker, and J.~Enns.
\newblock Highlighting {Interventions} and user differences: Informing adaptive
  information visualization support.
\newblock In {\em \acmchi}, p. 1835–1844, 2014.

\bibitem{Carrera2018TeachingWA}
C.~C. Carrera, J.~L.~S. P{\'e}rez, and J.~de~la Torre~Cantero.
\newblock Teaching with ar as a tool for relief visualization: usability and
  motivation study.
\newblock {\em International Research in Geographical and Environmental
  Education}, 27:69 -- 84, 2018.

\bibitem{chen_individual_2000}
C.~Chen.
\newblock Individual differences in a spatial-semantic virtual environment.
\newblock {\em Journal of the American Society for Information Science},
  51(6):529--542, 2000.

\bibitem{chen_spatial_1997}
C.~Chen and M.~Czerwinski.
\newblock Spatial {Ability} and {Visual} {Navigation}: an {Empirical} {Study}.
\newblock {\em Sability and visual navigation: an empirical study, New Review
  of Hypermedia and Multimedia}, 3(1):67--89, 1997. doi: {{%
10\hspace{.1pt}\discretionary{.}{%
}{.}\hspace{.4pt}1080\discretionary{/}{%
}{/}13614569708914684}}


\bibitem{cohen_individual_2007}
C.~Cohen and M.~Hegarty.
\newblock Individual differences in use of external visualizations to perform
  an internal visualization taks.
\newblock {\em Applied Cognitive Psychology}, 21:701--711, 2007. doi: {{%
10\hspace{.1pt}\discretionary{.}{%
}{.}\hspace{.4pt}1002\discretionary{/}{%
}{/}acp\hspace{.1pt}\discretionary{.}{%
}{.}\hspace{.4pt}1344}}


\bibitem{Cohen1969StatisticalPA}
J.~Cohen.
\newblock {\em Statistical Power Analysis for the Behavioral Sciences}.
\newblock Routledge, 1988.

\bibitem{collins1985self}
J.~L. Collins.
\newblock Self-efficacy and ability in achievement behavior.
\newblock In {\em Annual Meeting of the American Educational Research
  Association.}, 1982.

\bibitem{Colmane006928}
E.~Colman, M.~Wanat, H.~Goossens, S.~Tonkin-Crine, and S.~Anthierens.
\newblock Following the science? views from scientists on government advisory
  boards during the covid-19 pandemic: a qualitative interview study in five
  european countries.
\newblock {\em BMJ Global Health}, 6(9), 2021.

\bibitem{msoftcovid-19}
M.~P. Community.
\newblock Covid-19 data stories gallery, 2020.

\bibitem{conati_evaluating_2014}
C.~Conati, G.~Carenini, E.~Hoque, B.~Steichen, and D.~Toker.
\newblock Evaluating the {Impact} of {User} {Characteristics} and {Different}
  {Layouts} on an {Interactive} {Visualization} for {Decision} {Making}.
\newblock {\em Computer Graphics Forum}, 33(3):371--380, June 2014.

\bibitem{conati_eye_2014}
C.~Conati and D.~Toker.
\newblock Eye {Tracking} to {Understand} {User} {Differences} in
  {Visualization} {Processing} with {Highlighting} {Interventions}.
\newblock In {\em User {Modeling}, {Adaptation}, and {Personalization}}, vol.
  8538 of {\em {LNCS}}. Springer, 2014.

\bibitem{Downing2005TheEA}
R.~E. Downing, J.~L. Moore, and S.~W. Brown.
\newblock The effects and interaction of spatial visualization and domain
  expertise on information seeking.
\newblock {\em Comput. Hum. Behav.}, 21:195--209, 2005.

\bibitem{Dragicevic2016}
P.~Dragicevic and M.~Kaptein.
\newblock {\em Fair Statistical Communication in HCI}, pp. 291--330.
\newblock Springer International Publishing, Cham, 2016.

\bibitem{eurostat_2020}
Ec.europa.eu.
\newblock Tertiary education statistics, 2020.

\bibitem{ekstrom_manual_1976}
R.~Ekstrom, J.~French, H.~Harman, and D.~Derman.
\newblock Manual from {Kit} of {Factor}-{References} {Cognitive} {Tests}, 1976.

\bibitem{Elliott2021ADS}
M.~Elliott, C.~Nothelfer, C.~Xiong, and D.~A. Szafir.
\newblock A design space of vision science methods for visualization research.
\newblock {\em \tvcg}, 27:1117--1127, 2021.

\bibitem{Eppig2010ParasitePA}
C.~Eppig, C.~L. Fincher, and R.~Thornhill.
\newblock Parasite prevalence and the worldwide distribution of cognitive
  ability.
\newblock {\em Proceedings of the Royal Society B: Biological Sciences},
  277:3801 -- 3808, 2010.

\bibitem{FERGUSON20151}
A.~M. Ferguson, E.~A. Maloney, J.~Fugelsang, and E.~F. Risko.
\newblock On the relation between math and spatial ability: The case of math
  anxiety.
\newblock {\em Learning and Individual Differences}, 39:1--12, 2015.

\bibitem{few2007data}
S.~Few and P.~Edge.
\newblock Data visualization: past, present, and future.
\newblock {\em IBM Cognos Innovation Center}, 2007.

\bibitem{Figueredo2020TheBO}
A.~J. Figueredo, S.~C. Hertler, and M.~Pe{\~n}aherrera-Aguirre.
\newblock The biogeography of human diversity in cognitive ability.
\newblock {\em Evolutionary Psychological Science}, pp. 1--18, 2020.

\bibitem{Friendly2008ABH}
M.~Friendly.
\newblock A brief history of data visualization.
\newblock In {\em Handbook of Data Visualization}. Springer Handbooks Comp.
  Statistics., 2008.

\bibitem{GANLEY2011235}
C.~M. Ganley and M.~Vasilyeva.
\newblock Sex differences in the relation between math performance, spatial
  skills, and attitudes.
\newblock {\em Journal of Applied Developmental Psychology}, 32(4):235--242,
  2011.

\bibitem{Glynn2009ScienceMQ}
S.~M. Glynn, G.~Taasoobshirazi, and P.~Brickman.
\newblock Science motivation questionnaire: Construct validation with
  nonscience majors.
\newblock {\em Journal of Research in Science Teaching}, 46:127--146, 2009.

\bibitem{Greenland2016StatisticalTP}
S.~Greenland, S.~J. Senn, K.~J. Rothman, J.~B. Carlin, C.~L. Poole, S.~N.
  Goodman, and D.~G. Altman.
\newblock Statistical tests, p values, confidence intervals, and power: a guide
  to misinterpretations.
\newblock {\em European Journal of Epidemiology}, 31:337 -- 350, 2016.

\bibitem{Hall2020DesignBI}
K.~W. Hall, A.~J. Bradley, U.~Hinrichs, S.~Huron, J.~Wood, C.~M. Collins, and
  M.~S.~T. Carpendale.
\newblock Design by immersion: A transdisciplinary approach to problem-driven
  visualizations.
\newblock {\em \tvcg}, 26:109--118, 2020.

\bibitem{Hall2022ProfessionalDA}
K.~W. Hall, A.~Kouroupis, A.~Bezerianos, D.~A. Szafir, and C.~M. Collins.
\newblock Professional differences: A comparative study of visualization task
  performance and spatial ability across disciplines.
\newblock {\em \tvcg}, 28:654--664, 2022.

\bibitem{healy_2022}
Y.~Healy.
\newblock Circular barplot and distortion, 2022.

\bibitem{hegarty_diagrams_2004}
M.~Hegarty.
\newblock Diagrams in the {Mind} and in the {World}: {Relations} between
  {Internal} and {External} {Visualizations}.
\newblock In {\em Diagrammtic representation and inference}, pp. 1--13.
  Springer, 2004.

\bibitem{henshaw2018data}
A.~L. Henshaw and S.~R. Meinke.
\newblock Data analysis and data visualization as active learning in political
  science.
\newblock {\em Journal of Political Science Education}, 14(4):423--439, 2018.

\bibitem{Hoffman1975AnalysisIE}
K.~Hoffman.
\newblock {\em Analysis in Euclidean Space}.
\newblock Dover Publication, 1975.

\bibitem{jones_spatial_2008}
S.~Jones and G.~Burnett.
\newblock Spatial {Ability} and {Learning} to {Program}.
\newblock {\em Human Technology}, 4(1):47--61, 2008.

\bibitem{BootstrapJung}
K.~Jung, J.~Lee, V.~Gupta, and G.~Cho.
\newblock Comparison of bootstrap confidence interval methods for gsca using a
  monte carlo simulation.
\newblock {\em Frontiers in Psychology}, 10, 2019.

\bibitem{kellen_effects_2012}
V.~Kellen.
\newblock {\em The {Effects} of {Diagrams} and {Relational} {Complexity} on
  {User} {Performance} in {Conditional} {Probability} {Problems} in a
  {Non}-{Learning} {Context}}.
\newblock PhD thesis, DePaul University, College of Computing and Digital Media
  Dissertations, 2012.

\bibitem{kim_explaining_2017}
Y.-S. Kim, K.~Reinecke, and J.~Hullman.
\newblock Explaining the {Gap}: {Visualizing} {One}’s {Predictions}
  {Improves} {Recall} and {Comprehension} of {Data}.
\newblock In {\em \acmchi}, pp. 1375--1386, May 2017.

\bibitem{kimura_sex_2000}
D.~Kimura.
\newblock {\em Sex and {Cognition}}.
\newblock MIT press, 2000.

\bibitem{KirbyCollab}
R.~M. Kirby and M.~Meyer.
\newblock Visualization collaborations: What works and why.
\newblock {\em IEEE Computer Graphics and Applications}, 33(6):82--88, 2013.

\bibitem{lee_correlation_2019}
S.~Lee, B.~C. Kwon, J.~Yang, B.~C. Lee, and S.-H. Kim.
\newblock The {Correlation} between {Users}' {Cognitive} {Characteristics} and
  {Visualization} {Literacy}.
\newblock {\em Applied Sciences}, 9(3):488, Jan. 2019.

\bibitem{liu_mental_2010}
Z.~Liu and J.~Stasko.
\newblock Mental models, visual reasoning and interaction in information
  visualization: a top-down perspective.
\newblock {\em \tvcg}, 16(6):999--1008, Oct. 2010.

\bibitem{tabcovid-19}
T.~LLC.
\newblock Global covid-19 tracker, 2020.

\bibitem{luo2019user}
W.~Luo.
\newblock User choice of interactive data visualization format: The effects of
  cognitive style and spatial ability.
\newblock {\em Decision Support Systems}, 122:113061, 2019.

\bibitem{MicallefAssessing12}
L.~Micallef, P.~Dragicevic, and J.-D. Fekete.
\newblock Assessing the effect of visualizations on bayesian reasoning through
  crowdsourcing.
\newblock {\em \tvcg}, 18(12):2536--2545, 2012.

\bibitem{Michael1957TheDO}
W.~B. Michael, J.~P. Guilford, B.~Fruchter, and W.~S. Zimmerman.
\newblock The description of spatial-visualization abilities.
\newblock {\em Educational and Psychological Measurement}, 17:185 -- 199, 1957.

\bibitem{Miller1956TheMN}
G.~A. Miller.
\newblock The magical number seven plus or minus two: some limits on our
  capacity for processing information.
\newblock {\em Psychological review}, 63 2:81--97, 1956.

\bibitem{MIX2012197}
K.~S. Mix and Y.-L. Cheng.
\newblock Chapter 6 - the relation between space and math: Developmental and
  educational implications.
\newblock In J.~B. Benson, ed., {\em Advances in Child Development and
  Behavior}, vol.~42, pp. 197--243. JAI, 2012.

\bibitem{NestedModel2009}
T.~Munzner.
\newblock A nested model for visualization design and validation.
\newblock {\em \tvcg}, 15(6):921--928, 2009.

\bibitem{nachar2008mann}
N.~Nachar et~al.
\newblock The mann-whitney u: A test for assessing whether two independent
  samples come from the same distribution.
\newblock {\em Tutorials in quantitative Methods for Psychology}, 4(1):13--20,
  2008.

\bibitem{harvardCOVID-19}
H.~C.~A. Network.
\newblock Covid-19 dashboards: Examples from the civic analytics network, 2022.

\bibitem{whocovid}
W.~H. Organization.
\newblock Who coronavirus (covid-19) dashboard, 2022.

\bibitem{Orionearthscience}
N.~Orion, D.~Ben-Chaim, and Y.~Kali.
\newblock Relationship between earth-science education and spatial
  visualization.
\newblock {\em Journal of Geoscience Education}, 45(2):129--132, 1997.

\bibitem{ottley_improving_2016}
A.~Ottley, E.~M. Peck, L.~Harrison, D.~Afergan, C.~Ziemkiewicz, H.~Taylor,
  P.~Han, and R.~Chang.
\newblock Improving {Bayesian} {Reasoning}: {The} {Effects} of {Phrasing},
  {Visualization}, and {Spatial} {Ability}.
\newblock {\em \tvcg}, 22(1), Jan. 2016.

\bibitem{PalanProlific}
S.~Palan and C.~Schitter.
\newblock Prolific.ac—a subject pool for online experiments.
\newblock {\em Journal of Behavioral and Experimental Finance}, 17, 12 2017.

\bibitem{peck_data_2019}
E.~M. Peck, S.~E. Ayuso, and O.~El-Etr.
\newblock Data is {Personal}: attitudes and perception of data visualization in
  rural pennsylvania.
\newblock In {\em \acmchi}, pp. 1--12, 2019.

\bibitem{Peck3DModel}
E.~M. Peck, B.~F. Yuksel, L.~Harrison, A.~Ottley, and R.~Chang.
\newblock Towards a 3-dimensional model of individual cognitive differences:
  Position paper.
\newblock In {\em \acmchi}, BELIV '12. New York, NY, USA, 2012.

\bibitem{Puth2015OnTV}
M.-T. Puth, M.~Neuh{\"a}user, and G.~D. Ruxton.
\newblock On the variety of methods for calculating confidence intervals by
  bootstrapping.
\newblock {\em The Journal of animal ecology}, 84:892--7, 2015.

\bibitem{worldindata}
H.~Ritchie, E.~Mathieu, L.~Rodés-Guirao, C.~Appel, C.~Giattino,
  E.~Ortiz-Ospina, J.~Hasell, B.~Macdonald, D.~Beltekian, and M.~Roser.
\newblock Coronavirus pandemic (covid-19) - our world in data, 2022.

\bibitem{Rushton2005THIRTYYO}
J.~P. Rushton and A.~R. Jensen.
\newblock Thirty years of research on race differences in cognitive ability.
\newblock {\em Psychology, Public Policy and Law}, 11:235--294, 2005.

\bibitem{Ryan2000SelfdeterminationTA}
R.~M. Ryan and E.~L. Deci.
\newblock Self-determination theory and the facilitation of intrinsic
  motivation, social development, and well-being.
\newblock {\em The American psychologist}, 55:68--78, 2000.

\bibitem{Salthouse1990AgeAE}
T.~Salthouse, R.~L. Babcock, E.~Skovronek, D.~R.~D. Mitchell, and R.~Palmon.
\newblock Age and experience effects in spatial visualization.
\newblock {\em Developmental Psychology}, 26:128--136, 1990.

\bibitem{schooten_exploring_2009}
B.~v. Schooten, A.~Suinesiaputra, R.~J. v.~d. Geest, B.~v. Dijk, J.~Reiber,
  P.~Sloot, and E.~Zudilova-Seinstra.
\newblock Exploring individual user differences in the {2D}/{3D} interaction
  with medical image data.
\newblock {\em Virtual Reality}, 14(2):105--118, 2009.

\bibitem{DSC:7Scen2018}
M.~Sedlmair.
\newblock Design study contributions come in different guises: Seven guiding
  scenarios.
\newblock In {\em Proceedings of the Sixth Workshop on Beyond Time and Errors
  on Novel Evaluation Methods for Visualization}, BELIV '16, p. 152–161.
  Association for Computing Machinery, New York, NY, USA, 2016.

\bibitem{DSM2012}
M.~Sedlmair, M.~Meyer, and T.~Munzner.
\newblock Design study methodology: Reflections from the trenches and the
  stacks.
\newblock {\em \tvcg}, 18(12):2431--2440, 2012.

\bibitem{Shea2001ImportanceOA}
D.~L. Shea, D.~Lubinski, and C.~P. Benbow.
\newblock Importance of assessing spatial ability in intellectually talented
  young adolescents: A 20-year longitudinal study.
\newblock {\em Journal of Educational Psychology}, 93:604--614, 2001.

\bibitem{Shen2003GuidanceOE}
J.~Shen, E.~M. Reingold, and M.~Pomplun.
\newblock Guidance of eye movements during conjunctive visual search: the
  distractor-ratio effect.
\newblock {\em Canadian journal of experimental psychology}, 57:76--96, 2003.

\bibitem{BertiniBridge2015}
S.~Simon, S.~Mittelstädt, D.~A. Keim, and M.~Sedlmair.
\newblock {Bridging the Gap of Domain and Visualization Experts with a
  Liaison}.
\newblock In E.~Bertini, J.~Kennedy, and E.~Puppo, eds., {\em Eurographics
  Conference on Visualization (EuroVis) - Short Papers}. The Eurographics
  Association, 2015.

\bibitem{stanney_information_1995}
K.~Stanney and G.~Salvendy.
\newblock Information {Visualization}: {Assisting} low spatial inidviduals with
  information access tasks through the use of visual mediators.
\newblock {\em Ergonomics}, 38(6):1184--1198, 1995.

\bibitem{Tan2010TheCI}
S.~H. Tan and S.~B. Tan.
\newblock The correct interpretation of confidence intervals.
\newblock {\em Proceedings of Singapore Healthcare}, 19:276 -- 278, 2010.

\bibitem{tegarden1999business}
D.~P. Tegarden.
\newblock Business information visualization.
\newblock {\em Communications of the Association for Information Systems},
  1(1):4, 1999.

\bibitem{toker_gaze_2019}
D.~Toker, C.~Conati, and G.~Carenini.
\newblock Gaze {Analysis} of {User} {Characteristics} in {Magazine} style
  narrative visualizations.
\newblock {\em User Modeling, Adaptation, and Personalization}, 29:977--1011,
  2019.

\bibitem{tosto_why_2014}
M.~Tosto, K.~Hanscombe, C.~Haworth, O.~Davis, S.~Petrill, P.~Dale, S.~Malykh,
  R.~Plomin, and Y.~Kovas.
\newblock Why do spatial abilities predict mathematical performance?
\newblock {\em Wiley Developmental Science}, 17(3):462--470, Jan. 2014.

\bibitem{vanderplas_spatial_2016}
S.~VanderPlas and H.~Hofmann.
\newblock Spatial {Reasoning} and {Data} {Displays}.
\newblock {\em \tvcg}, 22(1):459--568, Jan. 2016.

\bibitem{velez_understanding_2005}
M.~Velez, D.~Silver, and M.~Tremaine.
\newblock Understanding visualization through spatial ability differences.
\newblock In {\em Vis '05}. IEEE, Minneapolis, MN, Oct. 2005.

\bibitem{vicente_assaying_1987}
K.~Vicente, B.~Hayes, and R.~Williges.
\newblock Assaying and isolating individual differences in searching a
  hierarchical file system.
\newblock {\em Human Factors: The Journal of the Human Factors and Ergonomics
  Society}, 29(3):349--359, 1987.

\bibitem{VidalFernandez}
M.~Vidal-Fernandez and D.~Yengin.
\newblock Underrepresentation of women in undergraduate economics degrees in
  europe: A comparison with stem and business.
\newblock {\em IZA - Institute of Labor Economics}, 2021.

\bibitem{Wai2009SpatialAF}
J.~Wai, D.~Lubinski, and C.~P. Benbow.
\newblock Spatial ability for stem domains: Aligning over 50 years of
  cumulative psychological knowledge solidifies its importance.
\newblock {\em Journal of Educational Psychology}, 101:817--835, 2009.

\bibitem{wenhong_user_2019}
L.~Wenhong.
\newblock User {Choice} of {Interactive} {Data} {Visualization} {Format}: {The}
  effects of {Cognitive} style and {Spatial} {Ability}.
\newblock {\em Decision Support Systems}, 122, 2019.

\bibitem{Wu2021UnderstandingDA}
K.~Wu, E.~Petersen, T.~Ahmad, D.~Burlinson, S.~Tanis, and D.~A. Szafir.
\newblock Understanding data accessibility for people with intellectual and
  developmental disabilities.
\newblock {\em \acmchi}, 2021.

\bibitem{Young2018TheCB}
C.~Young, S.~C. Levine, and K.~S. Mix.
\newblock The connection between spatial and mathematical ability across
  development.
\newblock {\em Frontiers in Psychology}, 9, 2018.

\bibitem{zheng2017data}
J.~G. Zheng.
\newblock Data visualization for business intelligence.
\newblock {\em Global business intelligence}, pp. 67--82, 2017.

\bibitem{ziemkiewicz_perceptions_2009}
C.~Ziemkiewicz and R.~Kosara.
\newblock Perceptions and individual differences in understanding visual
  metaphors.
\newblock {\em IEEE-VGTC Symposium on Visualization}, 28(3), 2009.

\bibitem{zinovyev2010data}
A.~Zinovyev.
\newblock Data visualization in political and social sciences.
\newblock {\em arXiv preprint arXiv:1008.1188}, 2010.

\end{thebibliography}
\end{document}